\documentclass[prd, reprint, showpacs, showkeys, nofootinbib,superscriptaddress]{revtex4-1}    

\usepackage{multirow}
\usepackage{amssymb,amsmath,bbm}
\usepackage{amsmath}
\usepackage{tensor}                                                    
\usepackage{graphicx}
\usepackage[usenames,dvipsnames]{color}                                
\usepackage{hyperref}
\usepackage{comment}                                                   
\usepackage{subfig}
\usepackage{enumerate}
\hypersetup{                 
    colorlinks=true,         
    linkcolor=blue,          
    citecolor=blue,           
    urlcolor=Violet          
}

\let\oldmarginpar\marginpar
\renewcommand\marginpar[1]{\oldmarginpar{\color{red}\raggedright\scriptsize #1}}
\newcommand{\pb}[2]{\ensuremath{\lf\{#1,#2 \rt\}}}
\newcommand{\mean}[1]{\ensuremath{\lf\langle #1 \rt\rangle }}

\newcommand{\diby}[2]{\ensuremath{\frac{\partial #1}{\partial #2}}}

\def\lf {\ensuremath{\left}}
\def\rt {\ensuremath{\right}}

\def\de {\ensuremath{ {\rm d} }}

\begin{document}

\title{\textbf{\Large Bouncing Unitary Cosmology\\II. Mini-Superspace  Phenomenology}}

\author{Sean Gryb}\email{sean.gryb@gmail.com}
\affiliation{{{\it Department of Philosophy}, University of Bristol}}
\affiliation{{{\it H. H. Wills Physics Laboratory}, University of Bristol}}
\author{Karim P. Y. Th\'ebault}\email{karim.thebault@bristol.ac.uk}
\affiliation{{{\it Department of Philosophy}, University of Bristol}}

\date{\today}


\begin{abstract}
A companion paper provides a proposal for cosmic singularity resolution based upon general features of a bouncing unitary cosmological model in the mini-superspace approximation. This paper analyses novel phenomenology that can be identified within particular solutions of that model. First, we justify our choice of particular solutions based upon a clearly articulated and observationally-motivated principle. Second, we demonstrate that the chosen solutions follow a classical mini-superspace cosmology before smoothly bouncing off the classically singular region. Third, and most significantly, we identify a `Rayleigh-scattering' limit for physically reasonable choices of parameters within which the solutions display effective behaviour that is insensitive to the details of rapidly oscillating Planck-scale physics. This effective physics is found to be compatible with an effective period of cosmic inflation well below the Planck scale. The detailed effective physics of this Rayleigh-scattering limit is provided via: i) an exact analytical treatment of the model in the de~Sitter limit; and ii) numerical solutions of the full model.
\end{abstract}

\maketitle
\tableofcontents


\section{Introduction}

Companion papers \cite{Gryb:2017a,Gryb:2018a} provide a proposal for cosmic singularity resolution based upon a quantum cosmology with a unitary bounce. This proposal is illustrated via a novel quantization of a mini-superspace model that leads to a finite, bouncing unitary cosmology in which there can be superpositions of the cosmological constant.\footnote{A mini-superspace model of the same form was derived some time ago in the context of unimodular gravity \cite{unruh:1989}. That earlier treatment did not include a detailed analysis of generic or specific cosmological solutions.  More problematically, the quantization did not include the self-adjoint representation of the Hamiltonian necessary to guarantee unitarity.} The resolution of classical singularities in a quantum theory is argued to be best analysed in terms of the generic finiteness of expectation values corresponding to classically divergent quantities. The bouncing unitary mini-superspace model is then shown to resolve the big bang singularity in this sense via the explicit construction of self-adjoint representations of the relevant observable algebra and Hamiltonian, which guarantees finiteness.  Whereas the object of the companion paper is to establish generic features of the general solutions -- such as singularity resolution -- the focus of the current paper is to consider physically salient features of particular solutions. Of particular importance will be the characteristics of a `cosmic beat' phenomenon and associated `bouncing envelope'. The cosmic beats can be interpreted in terms of the Planck-scale effects of the model and, under well-motivated constraints on the model parameters, are physically negligible relative to the effective envelope physics and are restricted to a `near-bounce' domain. In contrast, under these same constraints, the bouncing envelope persists into an effective regime of potential physical interest wherein the physics of the envelope displays universal phenomenology that is insensitive to the beat effects. Significantly, the limit in which the physics of the bouncing envelope is described on scales much larger than the Planck-sized beat physics is only available when superpositions of the cosmological constant are allowed. This behaviour thus constitutes a remarkable novel feature unique to bouncing unitary cosmologies. In what follows, we will first motivate physically relevant constraints on the general model parameters. We will then study, both analytically and numerically, the phenomenology of both the envelope and cosmic beats in the context of explicit solutions. Our ultimate aim is to connect the novel physics of the model with observational data, and a framework for further developments in that direction will be articulated in the final section.          

Perhaps the most difficult foundational problem of quantum cosmology is physically motivating a methodology for placing constraints on the form of the universal wavefunction. Since we are applying quantum theory to the entire universe, standard approaches for constraining the functional form of the wavefunction via reference to the actual boundary conditions of a lab based quantum system are clearly not going to be available. The approach we will take in this paper is to try and assume as little as we need to in order to extract physics from our model. In particular, we will avoid taking a stance on the vexed question of `interpreting' the universal wavefunction. Regardless of how it is  interpreted, in practice the functional form of the wavefunction used in our cosmological models is constrained by the information available to cosmologists. This observation motivates a general methodological principle for placing  constrains upon the form of the universal wavefunction. We call this principle: \textit{Epistemic Humility}. The principle demands that the conditions placed upon the wavefunction used in quantum cosmological models should involve the minimum possible assumption of information that we do not have.  As shall be argued in detail below, when combined with current observational data the principle of epistemic humility motivates a universal wavefunction that: i) is at all times Gaussian in a conserved momentum basis which includes the (square root of the) positive cosmological constant and the scalar field momentum; ii) has a time of minimal dispersion corresponding to the bounce time; and iii) becomes Gaussian and time-symmetric in a position basis during the current epoch. These requirements can be implement explicitly in the context of our model and restrict the form of the wavefunction to a family of particular solutions parametrised by a self-adjoint extension parameter that plays the role of a dimensionful reference scale. The parameters that determine the behaviour of the scalar field are left as free as possible, given the above constraints, so that they may be later fixed observationally in terms of a more concrete cosmological model.

Further motivations for the constraints imposed upon the model can be obtained from a physical interpretation of the dimensionful reference scale associated with the self-adjoint extensions. To this end, we will make use of the strong formal and physical analogy, introduced in the companion paper, that exists between bouncing unitary cosmology and a much studied $1/r^2$ effective model for three body atomic systems \cite{efimov:1970,gopalakrishnan:2006,ferlaino:2010}. The atomic analogy strongly suggests a connection between the requirement for a dimensionful self-adjoint extension parameter and the existence of a conformal anomaly. Fundamentally, the choice of self-adjoint extension is determined by the details of the micro-physics of the UV-completion of the effective system: i.e., the fundamental three body interactions. Such an interpretation is also natural in the cosmological model. Given this, the micro-physics of the underlying UV-completion of the mini-superspace limit of quantum general relativity should ultimately determine the value of self-adjoint extension parameter. The atomic analogy further motivates us to conceptualize the cosmological model in terms of a scattering experiment with a Gaussian wave-packet scattering off an effective potential produced by a microscopic state confined to the near-bounce region. The self-adjoint extension parameter can then be interpreted as a relative phase shift between the `in-going' and `out-going' energy eigenstates of the solution. The scattering experiment probes the UV-completion via these phase shifts.

Given that we do not, as yet, have access to any data regarding the Planck-scale UV-completion of the mini-superspace limit of quantum general relativity, epistemic humility motivates us to consider a choice of dynamics for the system that is as insensitive as possible to the Planck-scale physics, as encoded in the choice of self-adjoint extensions. This is precisely to operate with the minimum possible assumption of information that we do not have. Here we isolate a limit that contains an epistemically humble choice of self-adjoint extension and is simultaneously compatible with both the existence of a classically well-resolved positive cosmological constant and a physically reasonable large scalar field momentum. As is illustrated in detail in \S\ref{sub:generic sa ext}, this limit leads to universal behaviour of the self-adjoint extensions, as characterised by rapid cosmic beats associated with a large bouncing envelope. This behaviour is analogous to the existence of a \emph{Rayleigh-scattering} limit in the analogue system, where the wavelength of the incident photons is large compared with the size of the effective atomic system. It is a novel and highly advantageous feature of the model that such a Rayleigh limit exists: firstly, because this limit exhibits universal behaviour that relies on minimal assumptions regarding the underlying micro-physics of system; and secondly, because it is consistent with the foundational principles and empirical facts used to derive it.

The main results of the paper are as follows: i) an exact analytical treatment of the bouncing unitary cosmology model in the de~Sitter limit, where the scalar field momentum is taken to be vanishingly small; and ii) numerical solutions of the full model. Numerical evidence is provided in the full model for the existence of a Rayleigh limit, as described above, and for a semi-classical turnaround point in the dynamics of the scalar field which resembles an effective inflationary regime. The Rayleigh scattering limit implies that, as was already noted, the bounce can be observed to occur at a much lower energy scale than that of the Planck effects.

The contents of the paper are arranged as follows. Section \ref{sec:general} provides an overview of the mini-superspace model including its classical definition, the self-adjoint representation of the operator algebra, and the general solution. Section \ref{sec:methodological_foundations} provides further details regarding our methodological principle of epistemic humility and the restrictions that it motivates us to place on the universal wavefunction. Next, in Section \ref{sec:constraints}, we place specific constraints upon the general solutions based upon these foundational considerations. In particular, in  \ref{sub:state}, we will restrict the form of the wavefunction while in  \ref{sub:extension} we will fix the self-adjoint extension parameter. The main results of the paper are presented in Section \ref{sec:expl}. We begin, in Section \ref{sub:generic sa ext}, commenting briefly on the limit where the physics of any choice of self-adjoint extension becomes universal. Next, in Section \ref{sub:desitter}, we provide an exact analytical treatment of the model in the limit where the scalar field momentum is vanishingly small. Finally, in Section \ref{sub:numerical solution}, we study the physics of the remaining parameter space using numerical methods. In the conclusions we provide, in Section \ref{sec:concl}, an outline of the prospects for our model to be connected to inflationary cosmology.

\section{General Solutions}
\label{sec:general}
\subsection{Definition of Classical Model} 
\label{sub:definition_of_classical_model}


Here we will define the classical model explored in detail in the companion paper \cite{Gryb:2017a} and summarize the pertinent results for the current analysis. We consider an homogeneous and isotropic FLRW universe with zero spatial curvature ($k=0$) described by the scale factor, $a$, coupled to a massless free scalar field, $\phi$. In terms of these variables the space-time metric takes the form:
\begin{equation}\label{eq:metric}
  \de s^2 = - N(t)^2 \de t^2 + a(t)^2 \lf( \de x^2 + \de y^2 + \de z^2 \rt)\,
\end{equation}
with $\partial_i \phi = 0$.  The Hamiltonian, which is a proportional to a constraint, reduces to:
\begin{equation}\label{eq:miniham}
  H = N \lf[ - \frac \kappa {12 V_0 a} \pi_a^2 + \frac{1}{2V_0 a^3} \pi_\phi^2 + \frac{V_0 a^3}\kappa \Lambda \rt]\,,
\end{equation}
for $\pi_a$ and $\pi_\phi$ canonically conjugate to $a$ and $\phi$. In the above, $N$ is the \emph{lapse} function, $\Lambda$ is the cosmological constant, and $\kappa = 8 \pi G_N$, where $G_N$ is the Newton constant. These standard cosmological variables can be converted to more mathematically transparent ones for the analysis of this paper in terms of the canonical transformation
\begin{align}
  v &= \sqrt {\frac 2 3} a^{3} & \varphi &= \sqrt{\frac {3 \kappa}2} \phi \\
  \pi_v &= \sqrt{\frac 1 6} a^{-2}\pi_a & \pi_\varphi &= \sqrt{\frac 2 {3 \kappa}} \pi_\phi 
\end{align}
and the dimensionless lapse, $\tilde N$, and cosmological constant, $\tilde \Lambda$:
\begin{align}\label{eq:dimless N and Lambda}
  \tilde N &=  \sqrt{ \frac 3 2} \frac {\kappa \hbar^2 v N}{V_0}  & \tilde \Lambda &= \frac{ V_0^2} {\kappa^2\hbar^2} \Lambda \,.
\end{align}
For the moment, $\hbar$ represents some reference angular momentum scale for conveniently keeping track of units in the classical analysis. In terms of the above quantities, the Hamiltonian takes the form
\begin{align}
  H &= \tilde N \lf[ \frac{1} {2\hbar^2} \lf( - \pi_v^2 + \frac{\pi_\varphi^2}{v^2} \rt) + \tilde \Lambda \rt] \nonumber \\& =\tilde N \lf[ \frac 1 2 \eta^{AB} p_A p_B + \tilde \Lambda \rt]\,,\label{eq:v-phi Ham}
\end{align}
where $p_A = (\pi_v, \pi_\varphi)$ and $\eta^{AB}$ is the inverse of the Rindler metric $\eta_{AB}= \hbar^2 \text{diag}(-1,v^2)$, where we keep in mind that $v \in \mathbbm R^+$.

Hamiltonian's second equation for $\pi_\varphi$ implies that the $\varphi$ momentum is a conserved quantity which we will identify as $k_0$ in anticipation of the notion we will use in the quantum theory. This observation implies that the kinetic term for $\varphi$ can be integrated out to an effective $-1/v^2$ potential. This crucial observation links the mini-superspace theory to general systems with $-1/v^2$ potentials, for which there exists a vast phenomenologically rich literature. For a brief discussion of this literature and the link to our model, see \cite{Gryb:2017a}.

The remaining Hamilton equations can be integrated to give
\begin{align}
  v(\tau)^2 &= s^2 \lf[\lf(\frac{\tau - \tau_0}{\tau_s}\rt)^2 - 1 \rt] \\ \varphi(\tau) &= \varphi_\infty + \tanh^{-1}\lf( \frac {\tau_s} {\tau} \rt) \,,\label{eq:class solns}
\end{align}
where we have defined the quantities $\omega_0 = \sqrt{2\hbar^2 \tilde \Lambda}$, $s = |k_0/\omega_0|$, $\de \tau \equiv N \de t$, and $\tau_s = \hbar^2 k_0/\omega_0^2$. The time, $\tau_s$, represents the time of the classical singularity. The parameter $\omega_0$ has been defined in terms of the cosmological constant so that it asymptotes to the momentum conjugate to $v$ in the $\tau \gg 1$ limit, where the dynamics becomes approximately de~Sitter (i.e., dominated by the cosmological constant). The parameter $s$ plays the role of an `impact parameter' that specifies the relative momentum contained in the scalar field, and consequently sets the scale for when the dynamics of $\varphi$ will divert the solution away from the de~Sitter limit. The integration parameters $\varphi_\infty$ (representing the asymptotic value of $\varphi$ as $\tau \to \infty$) and $\tau_0$ can be shifted using the boost invariance of the Rindler metric and the time translational invariance of the theory. The conserved quantities $s$ and $\omega_0$ can be absorbed into a choice of units for space and time. The classical model thus involves no free parameters or external reference scale. The complete physics of the classical model us described by the shape of the unparameterized relational curve:
\begin{equation}\label{eq:class ri curve}
  v = s | \text{cosech} \lf( \varphi - \varphi_\infty \rt) |\,,
\end{equation}
which is given in Figure~\ref{fig:rep inv sols} below. Note that the Rinder horizon at $v=0$ represents the classically singular boundary of the configuration space. In the companion paper, we show explicitly that the classical solutions reach the boundary in finite proper time and are singular there according to the standard definitions.
\begin{figure}[h]
  \centering
  \includegraphics[width=\linewidth]{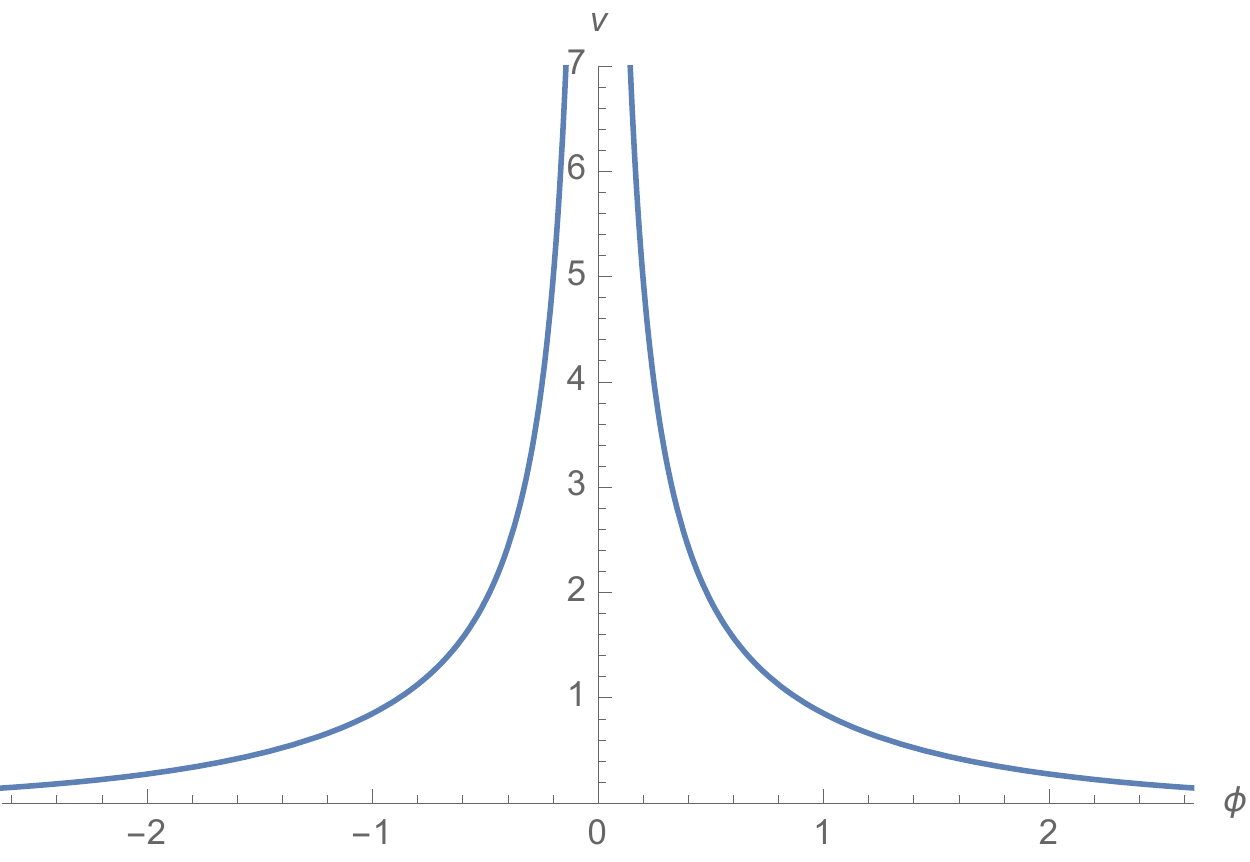}
  \caption{Classical solutions on $\mathcal C$ in units where $s = 1$ and $\varphi_\infty = 0$.} \label{fig:rep inv sols}
\end{figure}

\subsection{General Quantum Theory} 
\label{sub:quantum_theory}

As is outlined in detail in \cite{Gryb:2017a}, a self-adjoint representation of the operator algebra can be given in terms of the \emph{tortoise} coordinate
\begin{equation}
  \mu = \log v\,
\end{equation}
as
\begin{align}
  \hat\mu \Psi &= \mu \Psi & \hat\pi_\mu &= -i\hbar e^{-\mu} \diby{}{\mu}\lf( e^{\mu} \Psi \rt) \\
  \hat\varphi \Psi &= \mu \Psi & \hat\pi_\varphi &= -i\hbar \diby{\Psi}{\varphi}\,.
\end{align}
This field redefinition is necessary because the operator $\hat\pi_v$, conjugate to the volume, is not essentially self-adjoint, which is the case for any shift operator defined on $\mathbbm R^+$. The Schr\"odinger wavefunction, $\Psi$, is an element of the Hilbert space of square integrable functions $L(\mathbbm R^2, e^{2\mu})$ under the measure
\begin{equation}\label{eq:inner prod}
   \mean{\Phi,\Psi} = \int_{\mathbbm{R}^2} \de \varphi \de \mu\, e^{2\mu}\, \Phi^\dag \Psi\,.
\end{equation}
A functional basis for this Hilbert space is given by the span of the orthonormal eigenstates, $\psi_r$ and $\psi_k$, of the $\hat\pi_\mu$ and $\hat\pi_\varphi$ operators:
\begin{align}
  \psi_{r} &= \frac 1 { 2\pi \hbar }e^{-\tfrac i \hbar \mu r - \mu} & \psi_k &= \frac 1 { 2\pi \hbar } e^{-\tfrac i \hbar \varphi k}\,.
\end{align}

A natural quantization of the Hamiltonian, \eqref{eq:v-phi Ham}, in $v\varphi$-variables is given by
\begin{equation}
  \hat H = \frac 1 2 \lf( \diby{^2}{v^2} - \frac 1 {v^2} \diby{^2}{\phi^2}  \rt)\,.
\end{equation}
As outlined in detail in the companion paper \cite{Gryb:2017a}, relational quantization\footnote{See \cite{gryb:2011,gryb:2014,Gryb:2016a} for further motivation and applications of relational quantization.} of this model suggests a unitary evolution equation of the Schr\"odinger form:\footnote{Note that $t$ is consistent with its definition in \eqref{eq:metric} with $N=1$ and should be distinguished from $\de \tau = N \de t$ defined after \eqref{eq:class solns}.}
\begin{equation}\label{eq:seq}
  \hat H \Psi = i \hbar \diby{\Psi}{t}\,,
\end{equation}
where we interpret $t$ as a non-observable label whose only role in the theory is to order states of the universe at successive instants. The eigenstates,
\begin{equation}
  \hat H \Psi^\pm(v, \varphi) = \tilde \Lambda \tilde\Psi^\pm_\Lambda(v, \varphi)\,,
\end{equation}
of $\hat H$ can be computed by separation of variables, $\Psi_{\Lambda}^\pm (v, \varphi) = \psi_{\Lambda, k}(v) \nu_{k}^\pm (k)$, in terms of the the solutions of the wave equation
\begin{equation}
  \frac{\de^2}{\de \varphi^2} \nu_k^\pm = - \frac {k^2}{\hbar^2} \nu_k^\pm\,,
\end{equation}
for the $\varphi$-dependence, and Bessel's equation
\begin{equation}\label{eq:bessel}
  v \frac{\de }{\de v}\lf( v \frac {\de}{\de v}  \psi_{\Lambda,k} \rt) + \lf( 2 \tilde \Lambda v^2 + \frac{k^2}{\hbar^2} \rt) \psi_{\Lambda,k} = 0
\end{equation}
for the $v$-dependence.

The solutions of Bessel's equation are qualitatively different depending on the sign of $\Lambda$. The self-adjointness of $\hat H$ leads to a discrete `bound' spectrum for $\Lambda < 0$ in terms of the modified Bessel functions of the second kind, $K_{ik/\hbar}(\sqrt{2\tilde \Lambda} v)$, which decay like $e^{-v}$. For $\Lambda > 0$, self-adjointness allows for a continuous `bound' spectrum in terms of linear combinations of the Bessel functions of the first, $J_{ik/\hbar}(\sqrt{2\tilde \Lambda} v)$, and second kind, $Y_{ik/\hbar}(\sqrt{2\tilde \Lambda} v)$, whose norm under the inner product \eqref{eq:inner prod} behaves like a cosine function for large $v$. The general solutions is then given by
\begin{widetext}
\begin{equation}\label{eq:gen soln}
  \Psi(v,\varphi, t) = \frac 1 {\sqrt 2} \lf[  \sum_{n= -\infty}^\infty e^{i \tilde \Lambda_n t/\hbar} E_n \Psi^\text{bound}_{-\Lambda_n} + \int_{0}^\infty \de \tilde\Lambda\, e^{ -i \tilde\Lambda t/\hbar } E(\tilde \Lambda) \Psi^\text{unbound}_{\Lambda,\Lambda_\text{ref}}  \rt]\,,
\end{equation}
where
\begin{align}\label{eq:boundunbound}
  \Psi^\text{bound}_\Lambda(v, \varphi) &= \sqrt{\frac {2} {\pi^2 \hbar}} \int_{-\infty}^\infty \de k\, \lf( A \cos \lf( \tfrac {k\varphi} \hbar \rt) + B \sin \lf( \tfrac {k\varphi} \hbar \rt) \rt) \sqrt{ \frac { |\tilde \Lambda| \sinh \lf( \pi k/\hbar \rt)}k }  K_{i k/\hbar} \lf( \sqrt{2 |\tilde \Lambda |} v \rt)\\
  \Psi^{\text{unbound}}_{\Lambda, \Lambda_\text{ref}}(v,\varphi) &= \int_{-\infty}^\infty \de k\, \frac{ C \cos \lf( \tfrac {k\varphi}{\hbar} \rt) + D \sin \lf( \tfrac {k\varphi}{\hbar} \rt)}{\sqrt{2 \pi}\hbar\lf| \cosh \lf( \tfrac {\pi k}{2\hbar} + i \tfrac k {2\hbar} \log\lf[ \tfrac {\Lambda}{\Lambda_\text{ref}(k)} \rt] \rt)  \rt|} \text{\cal{Re}} \lf[  \lf( \frac {\Lambda} {\Lambda_\text{ref}(k)} \rt)^{-ik/2\hbar} \mathcal J_{ik/\hbar}(\sqrt{2 \tilde \Lambda} v) \rt]\,,
\end{align}\end{widetext}
The distributions $A(k)\de k/\hbar, B(k)\de k/\hbar, C(k) \de k/\hbar$, $D(k) \de k/\hbar$ and $E(\tilde \Lambda)\de \tilde \Lambda$ are normalized such that the integral of their norm squared over the appropriate integration range is equal to 1. This guarantees that
\begin{align}
  \mean{\Psi^\text{unbound}_{\Lambda_a,\Lambda_\text{ref}}, \Psi^\text{unbound}_{\Lambda_b,\Lambda_\text{ref}} } &= \delta(\tilde \Lambda_a - \tilde \Lambda_b)\\ \mean{\Psi^\text{bound}_{\Lambda_a}, \Psi^\text{bound}_{\Lambda_b} } &= \delta_{ab}\,.
\end{align}
Inspection of the explicit form of the eigenstates above reveals that those with negative $\Lambda$ have odd parity under $k$ while those with positive $\Lambda$ are even under $k$ inversions only when $\Lambda_\text{ref}(k)$ is also even (or, equivalently, $k$-independent).

The representations above give the self-adjoint extensions of $\hat H$, which are parameterized by arbitrary choices of $\Lambda_\text{ref}$. The parameter $\Lambda_\text{ref}$ enters the theory only through the combination $e^{i\theta}$, where
\begin{equation}\label{eq:theta}
  \theta = \frac k {2\hbar} \log \lf( \frac{\Lambda}{\Lambda_\text{ref}} \rt)\,
\end{equation}
has periodicity $2\pi$. This highlights the fact that the self-adjoint extensions form a $U(1)$ family where we must identify
\begin{equation}\label{eq:lambda periodicity}
  \log \Lambda_\text{ref} \to \log \Lambda_\text{ref} + \frac {2n\pi\hbar} k\,
\end{equation}
for $n \in \mathbbm Z$. This has been taken into account in the `bound' states through the definition of $\Lambda_n$:
\begin{equation}
  \Lambda_n \equiv e^{ 2n\pi\hbar/ k} \Lambda_\text{ref}\,.
\end{equation}
This periodicity implies that the limits $\Lambda_\text{ref} \to 0$ and $\Lambda_\text{ref} \to \infty$ are not well defined and, therefore, that neither of these limits can provide a preferred choice of self-adjoint extension.

The asymptotic properties of the Bessel functions can be used to infer a convenient physical interpretation of the self-adjoint extensions. In the limit where $v$ is large, the Bessel functions take the form
\begin{multline}\label{eq:asympt}
  \mathcal J_{ik/\hbar}\lf(\frac{v \omega}\hbar \rt) \approx \lf( \frac {2\hbar} {\pi v \omega} \rt)^{1/2} \lf[ \cos\lf( \frac{v \omega}\hbar - \frac{\pi i k}{2\hbar} - \pi/4 \rt) \rt. \\ \lf. + \mathcal O\lf(\tfrac 1 {v \omega} \rt) \rt]\,,
\end{multline}
where, inspired by the classical variables defined in \eqref{eq:class solns}, we define
\begin{equation}
	\omega = \sqrt{2 \tilde\Lambda } \hbar\,.
\end{equation}
Let us then assume that the wavefunction is peaked around the classical value $\omega_0$ and that large $v$ is defined by $v \ll \omega_0/\hbar$. Given this, we can us \eqref{eq:asympt} to show that the asymptotic form of the bound wavefunctions differ between $t \to \infty$ and $t\to -\infty$ by a phase factor, $\Delta$, equal to
\begin{equation}\label{eq:phase}
    \Delta \equiv \frac \pi 2 + 2 \arctan \lf[ \tanh \lf({\tfrac {\pi k}{2\hbar}}\rt) \tan \lf( \tfrac k {2\hbar} \log\lf( \tfrac \Lambda {\Lambda_\text{ref}} \rt) \rt) \rt]\,.
\end{equation}
This suggests that the bounce is analogous to a scattering process about a region characterised by a scattering length $v_s = \frac{\Delta \hbar}{2 \pi \omega_0}$ where new physics is expected to take over. Concrete predictions of our formalism are only trustworthy over scales much larger than $v_s$. The scattering length is an external reference scale that provides the physical interpretation of the Planck scale with our model.  

\section{Methodological Foundations} 
\label{sec:methodological_foundations}

Foundational questions regarding the interpretation of quantum mechanics evidently become more pressing at the cosmological scale \cite{Bell:2001}. The most difficult cluster of questions relate to how we should interpret the  wavefunction. Whilst one group of interpretations take the wavefunction to refer to the physical state of some quantum system \cite{Bohm,everett:1957}, another take the wavefunction to refer to knowledge or information \cite{heisenberg:1958,peierls:1991,caves:2002,fuchs:2002,HarriganSpekkens10}. Our methodology here is intended, as much as is possible, to be `interpretation neutral', in that we will assume its applicability does not depend upon the endorsement of a particular interpretation of the universal wavefunction. More positively, we will assume that, regardless of how it is interpreted, in practice the functional form of the wavefunction used in our cosmological models is constrained by the information available to cosmologists. For example, based upon what we know, we can reasonably rule out the universal wavefunction being in a definite negative eigenstate of the cosmological constant. In this sense, it is in practice unavoidable that the form of the universal wavefunction mirrors the knowledge available to cosmologists, even if ultimately one wishes the wavefunction to refer to something physical. Adopting this weakly epistemic approach allows us to steer clear of the vexed foundational problems of quantum cosmology, whilst simultaneously motivating a general methodological principle for constraining the form of the universal wavefunction. Our methodological principle is as follows: 
\begin{quote}
\textit{Epistemic Humility: conditions placed upon the universal wavefunction used in quantum cosmological models should involve the minimum possible assumption of information that we do not have.} 
\end{quote}
 We take it that such a principle will be acceptable to those who conceive of the universal wavefunction as potentially representing the physical state of the universe, just as much as it is to those who take it to represent knowledge or information. The methodological principle of epistemic humility will be crucial in constraining the form of \eqref{eq:gen soln} towards physically relevant particular solutions. In the following section, we will show how this principle can be applied towards motivating specific parameter choices. In the remainder of this section, we will apply epistemic humility to confront three aspects of the `cosmological measurement problem' in the context of our model following the useful division of \cite{Schlosshauer:2003zy}. 
  
The first aspect of the cosmological measurement problem is the most well-known: the problem of definite outcomes. If we apply the quantum formalism to the entire universe then we should expect the wavefunction to be in superposition of various observables. However, in our measurements of observables, in particular the cosmological constant, we only ever record definite values. Epistemic humility teaches us to assume as little as we can about the universal wavefunction whilst retaining consistency with our observations. Repeated measurements of the cosmological constant in our current epoch reveal an extremely small, definite value up to some resolution. This constrains our late-time wavefunction to be highly peaked about the measured value of $\Lambda$. It does not, however, necessitate that the wavefunction should be in a definite eigenstate of the cosmological constant. To impose such a condition would be precisely to assume information that we do not have unnecessarily. Moreover, allowing for the possibility for superpositions of the cosmological constant does not require us to commit to a multiverse cosmology, since we have left open the option for an epistemic interpretational of the wavefunction. Thus, we do not attempt to solve the problem of definite outcomes, but rather adopt an agnostic approach that can be reconciled with both a range of available interpretations and, most significantly, the available cosmological data.  

The second aspect of the cosmological measurement problem relates to the transition from quantum to classical dynamics in terms of a semi-classical regime. In the context of our model this amounts to motiving `early-time' and `late-time' constraints on the wavefunction. Given that we are dealing with a bouncing cosmology, we take `late-time' to correspond to large absolute times, $t\rightarrow \pm \infty$, and `early-time' to correspond to $t=0$.\footnote{In this context it would is most natural to situate a distinct `arrow of time' in each branch, both pointing away from the deep quantum bounce region. Such a `one past, two futures' interpretation would correspond to a quantum version of the `Janus universe' that has been studied in the context of scale invariant particle models \cite{barbour:2014}.} In general, epistemic humility leads us to constrain the wavefunction in a manner that preserves as much symmetry as possible  between the $t\rightarrow + \infty$ and $t\rightarrow -\infty$ branches. This is because the unitary evolution equation \eqref{eq:seq} is of the Schr\"odinger type and therefore invariant under the time reversal operation. Thus, any constraint that is not time symmetric will require a further specification of the time convention being used. More specifically, if we wish to choose a time of minimal dispersion that involves the least assumption of information that we do not have, then $t=0$ is uniquely selected. Furthermore, since there is no way to narrow down which of the two `branches' of the bouncing cosmology our observational data relate to, the choice most in keeping with epistemic humility is to assume a semi-classical regime in both $t\rightarrow \pm \infty$. Time symmetry combined with epistemic humility can thus motivate us to constrain the the early-time wavefunction to be minimally dispersive and the late-time wavefunction to be maximally semi-classical. The most difficult aspect of the problem is then to explicitly characterise this semi-classical regime.

To this end, epistemic humility motivates us to exploit a particular expression of the observables of the quantum formalism in terms of the generalized moments of the wavefunction (see \eqref{eq:moments defn} for the formal definition). The specification of these moments can be shown to give us a local trivialisation of the quantum phase space \cite{bojowald:2006b,brizuela:2014}. The expression of the quantum observables in terms of the moments of the wavefunction is particularly well-adapted since a necessary condition for semi-classicality is the existence of a particular canonical basis for the classical phase space in which the moments of higher order than two are vanishingly small \cite{brizuela:2014}. Furthermore, we can define the vanishing of the higher order moments without specifying preferred units by considering their size relative to the ratio of the variance to the mean of the wavefunction. This is equivalent to requiring that the non-Gaussianties of the wavefunction are very small (in a particular basis) and is thus in keeping with epistemic humility since it involves fixing the semi-classical regime in the minimally specific manner. We will give explicit details regarding the semi-classical conditions in the following section.

The third aspect of the cosmological measurement problem relates to the selection of a preferred basis. In our particular case, this problem manifests itself in the characterisation of the semi-classical regime in terms of the moment expansion. The problem is to identify the physically relevant canonical basis in which to impose the vanishing of the higher order moments: a wavefunction that is approximately Gaussian when expressed in one basis, can be highly non-Gaussian when expressed in another basis. Epistemic humility allows us to resolve this ambiguity by motivating the selection of basis that is minimally specific. In particular, by choosing a basis that is persevered by the dynamics we confirm to epistemic humility, since any other choice will require more detailed specification of information that we do not have. Given this, we should look for a basis specified in terms of self-adjoint operators which correspond to classical observables that are globally conserved along dynamical trajectories (i.e., commute with the Hamiltonian). This classical stability requirement then guarantees, via the Ehrenfest theorem, that the chosen operators will correspond to conserved quantities in a quantum mechanical sense.\footnote{It is important to note that what counts as a conserved quantity here depends on the choice of time function. See \cite[\S5]{Gryb:2016a} for detailed discussion of this point.} Crucially, the moments of the wavefunction, when expressed in this canonical basis of conserved quantities, will then be stable under the unitary time evolution. In the next section, we will explicitly show that such a stable basis exists and give the details for its construction.\footnote{Further evidence towards a hypothesis of Gaussianity in terms of conserved quantities can be motivated by environmental decoherence, where the interaction Hamiltonian of the system and the environment is small \cite{Schlosshauer:2003zy}. Crucially, the criterion of basis stability is common to our choice of preferred basis and to that achieved by decoherence theory via environmentally induced super-selection.}

\section{Constraining the Model}
\label{sec:constraints}
In the previous section, we proposed that conditions placed upon the universal wavefunction should involve the minimum possible assumption of information that we do not have and pointed out several concrete implications of this principle. In \ref{sub:state}, we will implement these requirements explicitly to restrict the form of the quantum state. In \ref{sub:extension}, we will offer further arguments to fix the self-adjoint extension parameter.     

\subsection{Form of the Wavefunction}
\label{sub:state}

The general solution for our model, (\ref{eq:gen soln}), is an arbitrary superposition of unbound positive $\Lambda$ and bound negative $\Lambda$ parts. The linearity of the Hamiltonian implies that bound and unbound wavefunctions can only have a significant effect upon each other when there is a significant overlap between them. As noted above, the model can be constrained by the experimental observation that the current universe is well-approximated by a semi-classical state with a definite positive $\Lambda$. When combined with linearity, this observation implies that the bound negative $\Lambda$ states cannot overlap with the unbound positive $\Lambda$ states in the semi-classical regime. This, in turn, restricts the bound part of the wavefunction to be confined to a region of configuration space where $v$ is much smaller than it is currently. Observational data thus cannot be used to further constrain the bound part of the wavefunction. Since we wish to make the minimum possible assumption of information that we do not have, we will therefore set the bound part of the wavefunction to vanish by requiring $A(k)=B(k)=0$ in (\ref{eq:boundunbound}). 

Given this restriction, \eqref{eq:gen soln} specifies the general wavefunction in terms of: i) the components of the wavefunction, $E(\Lambda)$, in a basis of eigenstates of the Hamiltonian; and ii) the components of the wavefunction, $C(k)$ and $D(k)$, in a basis of eigenstates of, $\hat\pi_\varphi$. The relative values of the coefficients $C(k)$ and $D(k)$ introduce a $k$-dependent phase shift between `in-going' and `out-going' $\hat\pi_\varphi$-eigenstates. As noted above, there is no way to narrow down which of the two `branches' of the bouncing cosmology our observational data relate to. Epistemic humility dictates that we should make the simplest choice compatible with our observations: that the phase difference between these two modes is an unobservable constant. This justifies the choice $D(k)=0$, which ensures that the wavefunction is symmetric in $\varphi$ about $t=0$. 

The remaining freedom in the form of the wavefunction is determined by fixing the functional form of $E(\Lambda)$ and $C(k)$. As noted above, these choices can be motivated by appeal to a late-time semi-classical regime as defined in terms of the moment expansion. Explicitly, using the notation of \cite{bojowald:2006b}, the generalized moments for a 4-dimensional classical phase space, $(q^1, q^2, p^1, p^2)$, are expressed as
\begin{multline}\label{eq:moments defn}
  G^{a_{k1}, a_{k2}}_{b_{k1},b_{k2}} = \mean{ \lf( \hat q^{k_{1}} - q^{k_{1}}\rt)^{a_{k1}}\lf( \hat q^{k_{2}} - q^{k_{2}}\rt)^{a_{k2}} \rt. \\ \lf. \times \lf( \hat p^{k_{1}} - p^{k_{1}}\rt)^{b_{k1}}\lf( \hat p^{k_{2}} - p^{k_{2}}\rt)^{b_{k2}} }_\text{Weyl}\,,
\end{multline}
for $ {{a_{ki}}, {b_{ki}} } = 0, 1, \hdots, \infty$. The Weyl subscript indicates completely symmetric ordering. The evolution equations and commutation relations for the system can be expressed in general terms as symplectic flow equations for these moments.\footnote{For the explicit statement of these equations, see \cite{bojowald:2006b}.} Our general solution \eqref{eq:gen soln} can be regarded as a solution to these flow equations expressed explicitly in terms of the momentum-space wavefunction, and can thus be understood as relating the pure momentum moments, $G^{0,0}_{b_{k1},b_{k2}}$, to all others. The specification of $E(\Lambda)$ and $C(k)$ can therefore be understood as a way of choosing these momentum moments such that the entire solution can then be determined from relevant commutation relations and flow equations.

To formulate an explicit proposal for a stable preferred basis we consider the Killing vectors of the classical configuration space. We make such a choice since the Killing vectors will give vector fields that will be preserved by the Hamiltonian. Although we lack a way of modeling an `environment' for our system, we note that this definition is at least consistent with the stability requirement of preferred bases resulting from environmentally induced super-selection using decoherence. Formulating a super-selection principle along these lines will be the subject of future investigations.

Since the configuration space of the model has the geometry of a 2D Rindler space, we know that it it must have 3 linearly independent local Killing vector fields. However, because of the presence of the Rinder horizon, only one of these is global. This implies that, in the quantum formalism, the operators corresponding to the non-global solutions to the Killing equation fail to be self-adjoint. We are therefore able to single out the boost generator as the unique global Killing vector field. The corresponding operator, $\hat\pi_\varphi$, is a natural choice for a stable basis for the moments of the wavefunction.

A candidate for the second component of the preferred basis can be derived from the asymptotic (i.e., large $v$) Killing vector field, $\partial/\partial v$. In the canonical language, for large $v$, the Killing equation translates into the approximate vanishing of the symplectic flow of $v$ under the flow of the Hamiltonian: $\pb{\pi_v}{H} \approx 0$. This is due to the fact that $H \propto \pi_v^2$ in this limit. This suggest a potential choice of a preferred basis in terms of eigenstates of $\hat\pi_v$. However, as was mentioned above and was discussed extensively in \cite{Gryb:2017a}, the operator $\hat\pi_v$ is not self-adjoint and, furthermore, is not globally conserved. The natural alternative is to use the self-adjointness of $\hat H$ and the asymptotic relation $H \propto \pi_v^2$ to motivate a preferred basis in terms of $\sqrt{\hat H}$. This choice is only unique up to a definition of operator ordering. However, we take such ordering ambiguity to only be significant for the deep UV physics of the model to which we are observationally ignorant. Choosing the preferred basis in terms of $\sqrt{\hat H}$ then suggests that we take Gaussian superpositions in $\omega$, which is proportional to the square root of the eigenvalue of $\hat H$. Asymptotically this will  give a Gaussian state in a basis of eigenstates of $\hat\pi_v$ and remain Gaussian throughout the entire evolution.

We are now able to formulate a completely unambiguous implementation of semi-classicality for our model in terms of the requirement that the wavefunction be Gaussian in the bases described above. Requiring that $\Lambda$ and $\pi_\varphi$ be well-resolved implies that the absolute value of the means of $E(\omega)$ and $C(k)$ must be much larger than the variances, otherwise the quantum mechanical uncertainty would make then indistinguishable from zero. Let us then define the uncertainty associated with $\omega$ to be $\sigma_\omega$ and the uncertainty associated with $k$ to be $\sigma_k$. We can now characterise the semi-classical forms of $E(\omega)$ and $C(k)$ as:
\begin{align} \label{eq:k and E wavefunction}
  \frac \omega \hbar E(\omega) &\approx \lf(  \frac{\hbar^2}{2 \pi \sigma_\omega^2}\rt)^{1/4} e^{- \frac { (\omega- \omega_0)^2}{4 \sigma_\omega^2} - \tfrac i \hbar (\omega- \omega_0) v_{0} } \\
  C(k) &\approx \lf(\frac{\hbar^2}{ 2\pi \sigma_k^2}\rt)^{1/4} e^{- \frac{(k- k_0)^2}{ 4 \sigma_k^2} - \tfrac i \hbar (k-k_0) \varphi_{\infty} } \,,\label{eq:k wavefunction}
\end{align}
where
\begin{align}\label{eq:moment limit}
  \omega_0 &\gg \sigma_\omega > 0 & |k_0| &\gg \sigma_k > 0\,.
\end{align}
and the density $E$ transforms such that $E(\Lambda) \to \frac \omega \hbar E(\omega)$.

In practice, because Gaussian states decay like the exponential of the square of the distance from the mean (in units of the variance), it follows that
\begin{align}
  \frac {\omega_0}{\sigma_\omega} &\ge 6 & \frac{|k_0|}{\sigma_k } & \ge 6\,
\end{align}
is sufficient to guarantee that the relevant uncertainty is reasonably small.\footnote{Quantitatively, this limit sets that $\omega_0$ and $k_0$ be different from zero to the six-sigma level -- or roughly 1 part in 1 million. The number $6$ was chosen somewhat arbitrarily and could easily be adapted to different applications. In a semi-classical limit valid at late times, one could interpret $\omega$ and $k$ as taking the definite values $\omega_0$ and $k_0$.}

For practical purposes, the standard form of a Gaussian, \eqref{eq:k and E wavefunction}, is not convenient for evaluating the $\omega$-integrals analytically. It will, therefore, be necessary to approximate \eqref{eq:k and E wavefunction} using a more convenient function that rapidly converges to a Gaussian in the limit we're working in. A convenient choice for such a function is:
\begin{multline}\label{eq:E approx}
  E(\omega) \approx \lf(\frac{\hbar^2 }{\sqrt{2 \pi} \sigma_\omega \omega}\rt)^{1/2}\lf( \frac{\omega}{\omega_0} \rt)^{ \omega_0^2/4\sigma_\omega^2} \\ \times \exp \lf\{  - \frac{2\omega_0^2}{\sigma_\omega^2} \lf[ \lf( \frac{\omega}{\omega_0} \rt)^2 - 1 \rt] - \frac i \hbar \lf( \omega - \omega_0 \rt)v_0 \rt\} \,.
\end{multline}
That this function gives a reasonable approximation to a Gaussian in the limit \eqref{eq:moment limit} is justified in Appendix~\ref{sub:regularization_of_gaussian}.

The parameters $v_0$ and $\varphi_\infty$ that appear in \eqref{eq:k and E wavefunction} and \eqref{eq:k wavefunction} represent the initial (i.e., t=0) value of $v$ and asymptotic value of $\varphi$ respectively. As was noted above, in the classical theory, these parameters can be shifted arbitrarily without loss of generality due to the time-translational and boost invariance of the theory. The same is true at the quantum level for $\varphi_\infty$. However, the interpretation of $v_0$ as the value of $v$ at $t=0$ no longer holds since quantum effects dominate the solution near $t=0$. Rather, as discussed in the previous section, we take $t=0$ to represent the time of minimum dispersion. Given that $v_0 = 0$ is the minimum possible value that this parameter can take, such a parametrisation corresponds to equating the time of minimum dispersion with the time when the expectation value of $v$ is at a minimum. Changing $v_0$ corresponds to shifting the time of minimum dispersion by an amount $v_0 \hbar^2/ \omega_0$, as can be seen by examining the asymptotic form of the wavefunctions. The virtue of the choice  $v_0 = 0$ is that it ensures that the $t<0$ and $t>0$ wavefunctions are indistinguishable and thus that time symmetry is persevered, as motivated by our epistemic humility principle. The choice of $\varphi_\infty$ can be motivated straightforwardly. The boost symmetry implies that this parameter reflects a genuine Killing direction of the classical configuration space, and thus we can choose $\varphi_\infty = 0$ without loss of generality.

The final consideration we shall make regarding constraining the form of the wavefunction relates to the physical role of units. We wish the restrictions we place upon the wavefunction to involve the least possible assumption of information we do not have. As noted in the previous section, the most natural way to specify units in this context is by reference to ratios. In particular, we can use the variances $\sigma_k$ and $\sigma_\omega$ to provide a set of units for the size of $k_0$ and $\omega_0$.  Consequently, the ratios $k_0/\sigma_k$ and $\omega_0/\sigma_\omega$ can be used to fix the $k$- and $\omega$-space wavefunctions without introducing any external reference scale. The final constraint on our model will, however, necessitate introducing an external reference scale. In particular, we need an additional external reference scale to fix the relative size of the Planck-scale effects in the universe. It is to this parameter that we now turn.

\subsection{Self-Adjoint Extension}
\label{sub:extension}

As was discussed at the end of \S\ref{sub:quantum_theory}, a physical interpretation of the self-adjoint extension parameter, $\Lambda_\text{ref}$, can be given in terms of the phase shift $\Delta$ through the relation explicitly expressed in \eqref{eq:phase}. This phase shift can be interpreted as a giving a particular scattering length in analogue atomic models. As was noted above in the companion paper \cite{Gryb:2017a}, a fascinating connection exists been between the physics of our cosmological model and the that of atomic 3-body systems. In particular, general solutions have a mathematical form that mirrors that of scattering of plane waves off bound atomic trimer states, the physics of which is described by an effective $1/r^2$ potential with $r$ playing an analogous role to $v$. In such a model, there is a scattering length, analogous to $v_s = 2 \pi \hbar \Delta/\omega_0$, determined by the micro-physics of the system, where the $1/r^2$ potential no longer accurately describes the system. 

The atomic analogy also leads us to connect the requirement for a dimensionful self-adjoint extension parameter to the existence of a conformal anomaly within the model. Formally, the anomaly breaks the fundamental scale-invariance of the $1/r^2$ potential by introducing a fundamental reference scale. In the context of the atomic model, this arises because the $1/r^2$ potential is only an effective description of the system. The choice of self-adjoint extension is determined by the details of the micro-physics of the UV-completion of this effective system. Such an interpretation is also natural to the cosmological model. One would reasonably assume that the homogeneous and isotropic approximation of quantum general relativity should break down at some energy scale: either because the assumption of homogeneity and isotropy break down or because quantized general relativity is found to be only an effective description of the physics in the early universe. Given this, the micro-physics of the underlying UV completion should ultimately determine the value of $\Lambda_\text{ref}$. Thus,  we have that role of $\Lambda_\text{ref}$ is to parametrize our ignorance of the UV completion of the model. Specifically, through the definition of $v_s$, $\Lambda_\text{ref}$ sets the scale where we expect the physics of the self-adjoint extensions to start to dominate the behaviour of the system. Since we have no way to know what the physics of the UV completion should be, it is best to regard $\Lambda_\text{ref}$ as a free parameter of our formalism that ideally would be fixed observationally.

Above we conveniently parametrized the $U(1)$ family of self-adjoint extensions using the reference scale $\Lambda_\text{ref}$, which enters the theory via the definition of the periodic variable $\theta$ in \eqref{eq:theta} as a way to give meaning to the units of $\Lambda$. Recalling the definition \eqref{eq:dimless N and Lambda} of the dimensionless cosmological constant, we see that the dimensionful quantity
\begin{equation}
  \frac{V_0^2}{\kappa^2 \hbar^2}
\end{equation}
sets the units of the cosmological constant. Since $\Lambda_\text{ref}$ enters the definition of $\theta$ exclusively through the ratio $\Lambda/\Lambda_\text{ref}$, these units can be completely absorbed into the definition of $\Lambda_\text{ref}$. Because $V_0$ can be rescaled by making an arbitrary choice of spatial units and $\kappa$ can be rescaled by changing the temporal units via its dependence on the dimensionless lapse in \eqref{eq:dimless N and Lambda}, the freedom to choose a self-adjoint extension by fixing $\Lambda_\text{ref}$ can be seen as a way of giving meaning to the value of $\hbar$. In other words, fixing $\Lambda_\text{ref}$ gives a scale with reference to which one can understand the relative size of quantum effects. This provides further support for an interpretation wherein the role of $\Lambda_\text{ref}$ is understood as demarcating unknown UV physics from that of our semi-classical universe. We will see these features play out explicitly in the analytically solvable model considered in \S\ref{sub:desitter} and that they persist in the full model of \S\ref{sub:numerical solution}.

In principle, direct observation of $\Lambda_\text{ref}$ could be achieved by measurement of the scattering length to determine $\Delta$. Although such measurements are possible in the analogue atomic systems, they are impractical from the perspective of cosmology because we only have knowledge of one `branch' of the bouncing cosmology. A reasonable hope is that, in more realistic cosmological models, where perturbative inhomogeneities are taken into account, $\Lambda_\text{ref}$ could have indirect empirical consequences in terms of the dynamics of the perturbations. For the moment, however, we must regard $\Lambda_\text{ref}$ as unconstrained by observation. Following our epistemic humility principle we can look for a parameter choice that is minimally specific. In particular, we can look to fix $\Lambda_\text{ref}$ without introducing any new parameters. Since the limits $\Lambda_\text{ref} \to 0$ and $\Lambda_\text{ref} \to \infty$ are not well defined, one natural thing to do is to set its value to the semi-classical value of the cosmological constant via:
\begin{equation}\label{eq:Lambda ref choice}
  \Lambda_\text{ref} = \frac{V_0^2}{\kappa^2 \hbar^2} \frac {\omega_0^2}{2\hbar^2},
\end{equation}
where we have used the full definition of $\Lambda$ in terms of all the parameters of the mini-superspace action. This achieves our goal of fixing $\Lambda_\text{ref}$ without introducing any new parameters. We will see in \S\ref{sub:generic sa ext} that this choice can also be motivated by requiring universality in a limit where $k_0/\hbar$ is large.

Combining the choice \eqref{eq:Lambda ref choice} with the restrictions of \ref{sec:methodological_foundations} and \ref{sub:state} we get as our general solution:
\begin{widetext}
\begin{equation}\label{eq:Psi full}
  \Psi(v, \varphi, t) = \sqrt{\frac 2 \pi}  \int_0^\infty \frac{\de k\, \de \omega\, \omega}{\hbar^3} \frac{e^{i \omega^2 t/ 2 \hbar^3} E(\omega) C(k) \cos \lf( \frac {k \varphi}{\hbar} \rt)}{\lf| \cosh \lf( \tfrac {\pi k}{2\hbar} + i \tfrac k {\hbar} \log\lf[ \tfrac {\omega}{\omega_0} \rt] \rt)  \rt|} \text{Re}\lf[ \lf( \frac{\omega}{\omega_0} \rt)^{-ik/\hbar} \mathcal J_{ik/\hbar}\lf(\tfrac {\omega v}\hbar\rt) \rt]\,.
\end{equation}\end{widetext}
with $E(\omega)$ and C(k) given by \eqref{eq:E approx} and \eqref{eq:k wavefunction} respectively.\footnote{Note that the choices $A = B = D = 0$ require a slight renormalization of the wavefunction and that $D=0$ implies that the wavefunction is even in $k$, allowing it to be written more conveniently in terms of an integral between $0$ and $\infty$.} The next section will study the properties of this general solution both analytically and numerically.

\section{Explicit Solutions}
\label{sec:expl}

Our main task in this section will be to study the physics of the model presented above via two independent investigations of the general character of the solutions \eqref{eq:Psi full}. Section \ref{sub:desitter} will be devoted to an exact analytical treatment of the model in the limit where the scalar field momentum is vanishingly small. In this limit, the late (i.e. large absolute time) semi-classical regime will be dominated by de~Sitter-like behaviour where the cosmological constant dominates the dynamics. The conditions \eqref{eq:moment limit} will be  shown analytically to imply the phenomena of cosmic beats and the bouncing envelope discussed in the introduction. Our analytic results show that the relative size of the bouncing envelope to that of the beats is determined by the ratio $\omega_0/\sigma_\omega$, which is the only independent parameter in the model. We interpret the beat phenomenon as a Planck-scale effect originating from the physics of a generic self-adjoint extension. We interpret the the limit $\omega_0/\sigma_\omega \gg 1$ as an analogue of Rayleigh scattering, where the Planck-scale effects remain small compared to the physics of the semi-classical envelope. In \S\ref{sub:numerical solution} we study the physics of the remaining parameter space using numerical methods. It is shown that the qualitative features of the Rayleigh scattering, which were described analytically in the de~Sitter limit, persist in the numerical solutions when the scalar field momentum is turned on, even whilst the approximation techniques used to gain an analytic understanding of these phenomenon break down. Numerical evidence is provided for a semi-classical turnaround point in the dynamics of the scalar field which resembles an effective inflationary epoch.

\subsection{Generic Self-Adjoint Extension Behaviour} 
\label{sub:generic sa ext}

Before investigating the detailed physics of our model, we comment briefly on a limit where the physics of any choice of self-adjoint extension becomes universal. This provides additional support both for the choice \eqref{eq:Lambda ref choice} and for our claims above regarding universality. The existence of the relevant limit relies on the compactness of the $U(1)$ group that parameterizes the space of self-adjoint extensions. This compactness leads to the periodicity, \eqref{eq:lambda periodicity}, previously discussed of the representation in terms of $\Lambda_\text{ref}$. The conditions that $k_0$ is large in units of $\hbar$ and well-resolved according to \eqref{eq:moment limit} jointly imply that
\begin{equation}
  e^{2\pi \hbar/k} \approx 1 + \frac{2 \pi \hbar}{k_0}\,.
\end{equation}
The periodicity in $\Lambda$ further implies that, for any choice of $\omega_\text{ref} = \sqrt{2 \Lambda_\text{ref}}$, there is an equivalent choice within a range
\begin{equation}\label{eq:omega range}
  e^{\pi \hbar/|k_0|} \omega_0
\end{equation}
of $\omega_0$. Thus, if we restrict to Gaussians in $\omega$ that satisfy the additional condition
\begin{equation}\label{eq:inflation limit}
  \frac {|k_0|} \hbar \gg \frac {\omega_0}{\sigma_\omega}\,,
\end{equation}
the $\omega_0$-periodicity \eqref{eq:omega range} becomes vanishingly small. This means that, for any choice of self-adjoint extension in terms of a reference scale $\omega_\text{ref}$, there is an equivalent choice imperceptibly close to $\omega_0$. All choices of self-adjoint extension are, therefore, equivalent to the choice \eqref{eq:Lambda ref choice} in the limit \eqref{eq:inflation limit}.


\subsection{de Sitter Limit}
\label{sub:desitter}

The de~Sitter limit is that in which the magnitude of the momentum of the scalar field is taken to be vanishingly small relative to the cosmological constant as measured in units of the widths of their distributions. Formally this is given by the condition: 
\begin{equation}\label{eq:dS limit}
 \frac{\omega_{0}/\sigma_\omega}{|k_0|/\sigma_k} \gg1\,.
\end{equation}
In this limit, a well-resolved Gaussian state is expect to be strongly peaked, in $v$-space, around the classical de~Sitter solution
\begin{equation}
  v(t) = \frac{\omega_0 |t|}{\hbar^2}\,.
\end{equation}
The absolute value indicates a gluing of the independent in- and out-going solutions at the classical singularity at $t=0$. Near the bounce at $t=0$, the quantum solution is expected to develop non-Gaussianities in $v$-space. These non-Gaussianities will prevent the expectation value of $v$ from attaining the singular value $\mean{v} = 0$. To model the quantum behaviour in the limit \eqref{eq:dS limit}, we take $\sigma_k/\hbar \gg 1$ while setting $k_0 = 0$. In such a limit, the $k$-space Gaussian can be effectively modelled by a $\delta$-function such that $C(k) = \delta\lf(\tfrac k \hbar\rt)$. The $k_0 = 0$ limit is ill-defined as a limit of the general choice of self-adjoint extension made thus far. However, a well-defined general solution, in this limit, can be conveniently expressed in terms of an arbitrary phase $\alpha$ between Bessel functions of the first, $\mathcal J_0$, and second, $\mathcal Y_0$ kind:\footnote{Although the $\mathcal Y_0$ are divergent at $v = 0$ they are nevertheless integrable under our measure. We need to introduce them here since when $k=0$, $\mathcal J_0$ is real, and so the imaginary components of $\mathcal J_0$ do not provide an independent solution of Bessel's equation.}
\begin{multline} \label{eq:ds wavefunction}
  \Psi(v,t) = \int_0^\infty \frac{\de \omega \omega} {\hbar^2} e^{i \omega^2 t/\hbar^3 } E(\omega/\hbar) \lf( \cos \tfrac \alpha 2 \mathcal J_0 \lf( \tfrac {\omega v}{\hbar} \rt) \rt. \\ \lf. - \sin \tfrac \alpha 2 \mathcal Y_0 \lf( \tfrac {\omega v}{\hbar} \rt) \rt)\,.
\end{multline}
The parameter $\alpha$ can be related to the asymptotic phase shift $\Delta$ using the large $v\omega$ expansions of the Bessel functions of order zero:
\begin{align}
    \mathcal{J}_{0}(\omega v/\hbar) &= \sqrt{\frac{2\hbar}{\pi \omega v}} \lf[ \cos \lf( \omega v/\hbar - \pi/4 \rt) + \mathcal O((\omega v/\hbar)^{-1}) \rt] \nonumber \\  \mathcal{Y}_{0}(\omega v/\hbar) &= \sqrt{\frac{2\hbar}{\pi \omega v}}\lf[ \sin \lf( \omega v/\hbar - \pi/4 \rt) + \mathcal O((\omega v/\hbar)^{-1}) \rt] \,.\label{eq:bessel 0 expansion}
\end{align}
Superposition of in- and out-going eigenstates in this limit leads to a phase shift of the form
\begin{equation}
  \Delta = \frac \pi 2 - \alpha\,.
\end{equation}

Comparison between this expression and \eqref{eq:phase} makes clear that the de~Sitter limit has the significant feature that the representations of the full $U(1)$ family of self-adjoint extensions do not require the introduction of a privileged reference scale such as $\Lambda_\text{ref}$. The classical conformal invariance discussed in \S\ref{sub:definition_of_classical_model} can thus be retained quantum mechanically, resulting in the absence of a conformal anomaly. An alternative but complementary way of seeing this is to notice that the full dependence of the theory on the dimensionful parameter $\hbar$ can be removed by the field redefinition $\bar\omega = \frac \omega \hbar$ followed by the time reparametrization $\bar t = \hbar t$. This illustrates that the only  free parameter of the theory that can play the role of an external reference scale in this limit is given by the large dimensionless ratio $\omega_0/\sigma_\omega$. To highlight this, the explicit $\hbar$ dependence has been omitted in the remainder of this sub-section.

\begin{figure}
  \subfloat[$v |\Psi|^2$ for $\omega_0/\sigma_\omega = 10$]{
    \includegraphics[width=\linewidth]{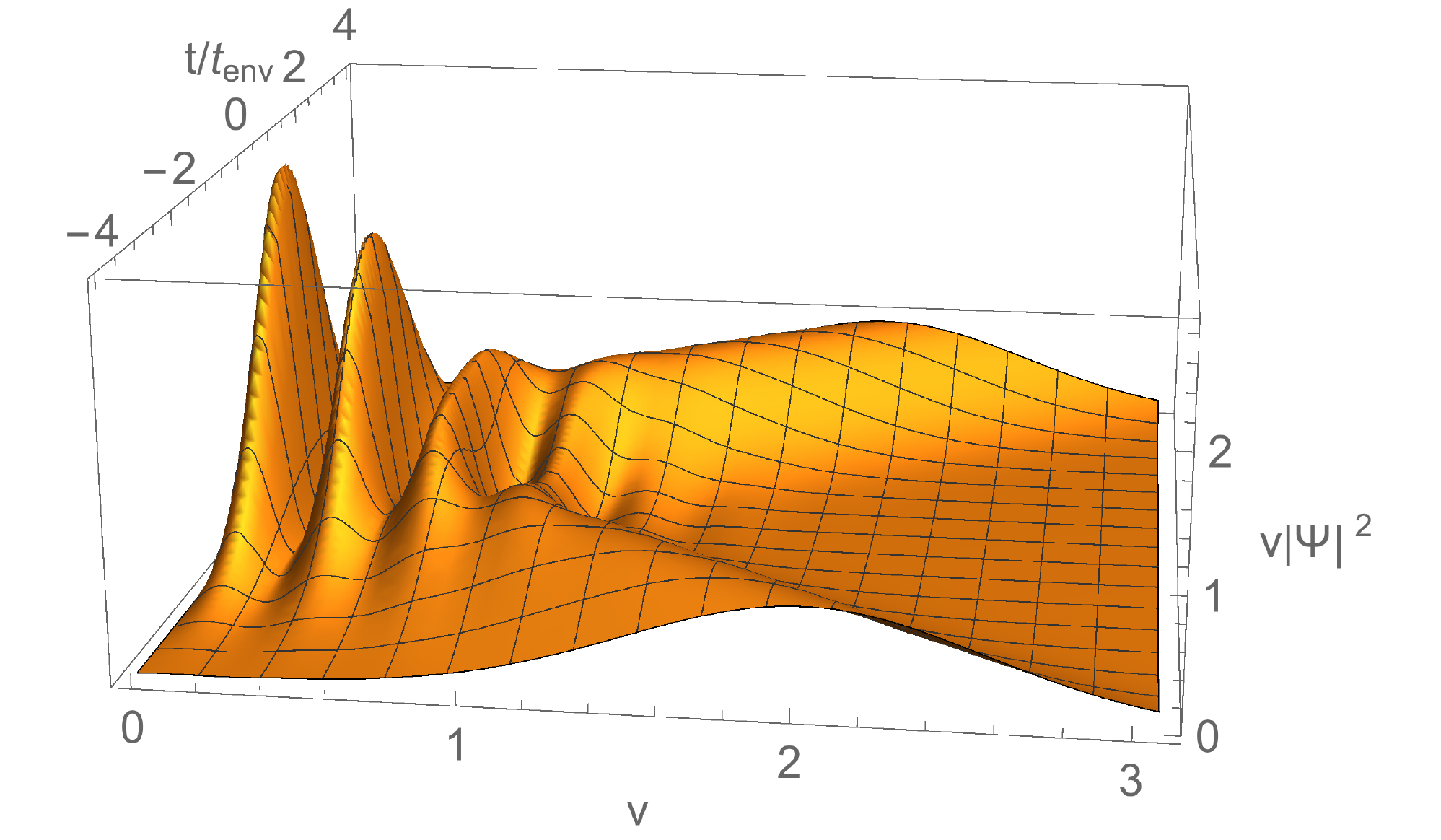}
    \label{fig:ds_omega_4}
  }\\
  \subfloat[$v |\Psi|^2$ for $\omega_0/\sigma_\omega = 15$]{
    \includegraphics[width=\linewidth]{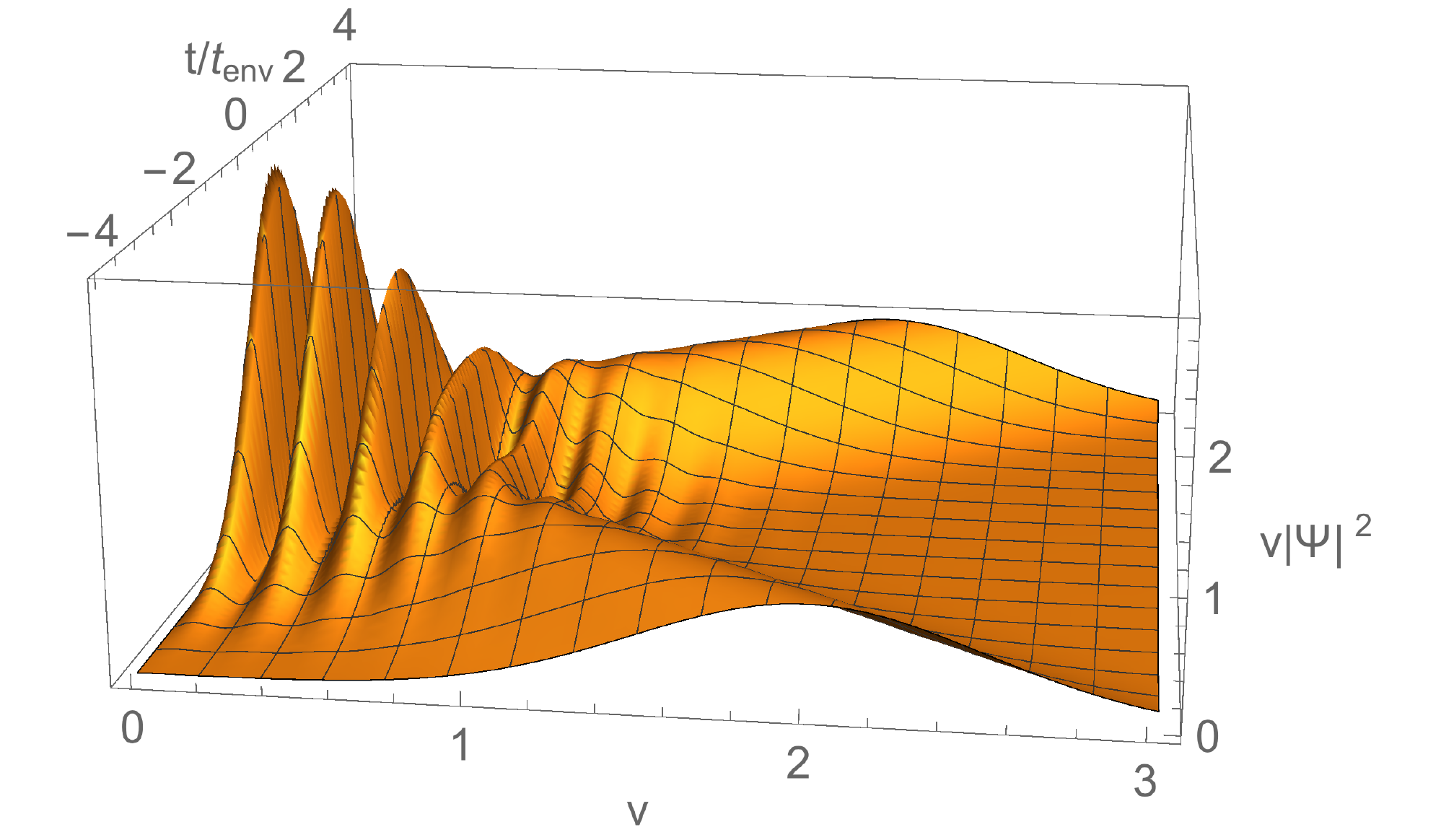}
    \label{fig:ds_omega_6}
  }
  \caption{\label{fig:psi_sq_ds} Time development of the exact Born amplitude for $\Psi(v,t)$ in the de~Sitter limit. Notice that $\omega_0/\sigma_\omega$ roughly sets the number of oscillations in the bounce region. Temporal units are set by the characteristic timescale set by the envelope size: $t_\text{env} \equiv 1/2 \sigma_\omega \omega_0$.}
\end{figure}

Use of the approximate Gaussian function given by \eqref{eq:E approx} for $E(\omega)$ in \eqref{eq:ds wavefunction} allows for explicit evaluation of the $\omega$-integrals in terms of analytic functions. The result of these integrations yields:
\begin{multline}\label{dssolution}
  \Psi(v,t) = \frac{ N_\text{dS} }{ 2 a^m }\lf[ \cos \tfrac \alpha 2 \Gamma\lf( \tfrac m 2 \rt) L_{- m/2} \lf( -\frac{ v^2 }{4a^2} \rt) \rt. \\ \lf. + \sin \tfrac \alpha 2\, G_{2,3}^{2,1}\left(\frac{v^2}{4 a^2}\lf|
\begin{array}{c}
 1-\frac{m}{2},-\frac{1}{2} \\
 0,-\frac{1}{2} \\
\end{array}\rt.
\right)\rt]\,,
\end{multline}
where we have introduced the Laguerre polynomials, $L_n(x)$, and the Meijer G-Function, $G_{i,j}^{k,l}\left( x \lf|\hdots \rt.\rt)$ and also defined the parameters 
\begin{align}
  N_\text{dS} &= \lf( \frac{1}{2\pi \sigma_\omega^2 } \rt)^{1/4} e^{\omega_0^2/8 \sigma_\omega^2} \omega_0^{3/2 - m}\nonumber\\
  a^2 &= \frac{1}{8 \sigma_\omega^2} - \frac{it}{2} \nonumber \\
  m &= \omega_0^2/4 \sigma_\omega^2 + \frac 3 2\,.\label{eq:ds parameters}
\end{align}
Plotting the expression \eqref{dssolution} for $\alpha = 0$ at the bounce time $t=0$ for moderately sized values of $\omega_0/\sigma_\omega$ (i.e., between 10 and 15) reveals the existence of a characteristic beat phenomenon in the overlap regions between in- and out-going solutions. Moreover, the value of $\omega_0/\sigma_\omega$ can be observed to be roughly inversely proportional to the wavelength of these beats. These features are illustrated in the plots of FIG.~\ref{fig:psi_sq_ds}.

A quantitative basis for the above qualitative observations can be provided by studying the analytic structure of both the beat phenomenon and the near-bounce wavefunction. This can be achieved, to a high degree of approximation, by appealing to the semi-classical constraints leading to the condition $\omega/\sigma_\omega \gg 1$. This condition tells us that the $\omega$-space wavefunction has most of its support in the region where $\omega \sim \omega_0$. Near the bounce, the expectation value of $v$ should be at a minimum. Since the wavefunction will have support in a region roughly the size of the variance of $v$, we posit that $\mean{v} \sim \sigma_v$ at $t = 0$. Moreover, as was previously argued, the operators $\hat v$ and $\hat \omega$ are approximately canonically conjugate in the $|t| \to \infty$ regimes so that $\sigma_v \sim 1/\sigma_\omega$ in this limit. It is, therefore, plausible to posit that this same relation should hold whenever the eigenstates of the wavefunction are well approximated by the large $v\omega$ expansion of the Bessel functions. Under this assumption, we find that, at the bounce, the following relation should hold:
\begin{equation}
  v\omega \sim \frac {\omega_0}{\sigma_\omega} \gg 1\,,
\end{equation}
so that use of the relation $\sigma_v \sim 1/\sigma_\omega$ is self-consistent. It is then possible to use the expansions \eqref{eq:bessel 0 expansion} to re-write the eigenstates of \eqref{eq:ds wavefunction} as:
\begin{multline}
  \cos \tfrac \alpha 2  \mathcal{J}_0 \lf(\omega v \rt) + \sin \tfrac \alpha 2 \mathcal{Y}_0 \lf(\omega v \rt) = \sqrt{\frac{2}{\pi \omega v}} \lf[ \cos \lf( \omega v - \Delta/2 \rt) \rt. \\ \lf.+ \mathcal O((\omega v)^{-1}) \rt]\,.
\end{multline}

Using the standard expression for a Gaussian in $E(\omega)$, we find that the wavefunction is given, to a very good approximation, by
\begin{multline}
  \psi(v,t) \approx \sum_\pm \lf( \frac 1 { 2^3 \pi^3 \sigma_\omega^2 }\rt)^{1/4} \int_{-\infty}^\infty \de \omega \exp \lf\{  - \frac{ (\omega \mp \omega_0)^2 }{4 \sigma_\omega^2} \rt. \\ \lf. + i \lf[  \omega v - \frac{\omega^2 t}2 \mp \frac \Delta 2 \rt]  \rt\}\,.
\end{multline}
The physics of the model is, therefore, well-described by a superposition of two Gaussian states following in- and out-going classical solutions that interfere near the bounce. The $\omega$-space integral can be evaluated using a variety of standard techniques. The result is
\begin{equation}\label{eq:ds approx wavefunction}
  \Psi(v,t) = \sum_\pm \mathcal N^\pm A^\pm e^{iS^\pm}\,,
\end{equation}
where
\begin{align}
  \mathcal N &\equiv \mathcal N^\pm =\lf( \frac 2 \pi \rt)^{1/4} \sqrt{\frac{ \sigma_\omega }{ 1 +  2i \sigma_\omega^2 t }} \nonumber\\
  A^\pm &= \exp \lf\{ - \frac{ \sigma_\omega^2 \lf(  v \mp \omega_0 t  \rt)^2 }{1 +4 \sigma_\omega^4 t^2 } \rt\} \nonumber\\
  S^\pm &= \frac{\pm \omega_0 v - \frac {\omega_0^2t}{2} + 2 \sigma_\omega^2 v^2 t }{ 1 + 4 \sigma_\omega^4 t^2  }\mp \Delta/2\,.\label{eq:ds phase}
\end{align}
The phases of the in-going, `$+$', and out-going, `$-$', states are given, to first order in quantum corrections, by the two independent solutions to the Hamilton--Jacobi equation for a de~Sitter universe shifted by a total phase of $\Delta$. The amplitudes are that of diffusing Gaussian wave-packets peaked on the classical histories and having minimum dispersion at $t=0$. A comparison can be made between the approximate wavefunction of \eqref{eq:ds approx wavefunction} and the exact one of \eqref{dssolution}. The near-bounce Born amplitudes of the exact and approximate wavefunctions are found to agree to better than two percent for $\omega_0/\sigma_\omega \geq 5$ and one percent for $\omega_0/\sigma_\omega \geq 10$.

Physical features of these solutions can be highlighted by computing the Born amplitude of the wavefunction in terms of the Guassian amplitudes, $A^\pm$, and the various parameters of the theory. From \eqref{eq:ds approx wavefunction} and the definitions \eqref{eq:ds phase}, we immediately obtain
\begin{multline}\label{eq:dsbeats}
  \lf| \Psi(v,t)  \rt|^2 = |\mathcal N|^2\lf[ (A^+)^2 + (A^-)^2 \rt. \\ \lf. + 2 A^+ A^- \cos\lf( \frac {2 \omega_0 v - \Delta}{1 + 4 \sigma_\omega^4 t^2 }  \rt) \rt]\,.
\end{multline} 
The first two terms represent dispersive Gaussian envelopes for the in- and out-going wave-packets, while the last term is an interference term representing rapid oscillations, or \emph{beats}, where the two envelopes overlap. These features of the solutions define two distinct length scales: the characteristic size of the envelope (ignoring dispersion effects), $v_\text{env} = 1/\sigma_\omega$, and the characteristic size of the beats, $v_\text{beat} = 1/\omega_0$. Note that, given our choice of parameters, the beat size is roughly equal to the scattering length, $v_s$, defined earlier. This lends further support for the interpretation of the beats in terms of the micro-physics of the underlying UV physics of the model. The condition $\omega_0/\sigma_\omega \gg 1$ implies that the beat phenomenon occurs on a much smaller length scale than the physics of the envelope, and, therefore, that Planck-scale effects are negligible on these scales. Moreover, the observation that $v_\text{env} \gg v_\text{beat}$ justifies our use of the terminology of `Rayleigh scattering' given the atomic analogy.

We can quantify deviations from classicality by explicitly computing the mean and variance of the wavefunction in terms of $v$. Because of the rapid oscillations of the beats in $v$, the first few moments, which are integrals over $v$, will be relatively insensitive to the detailed beat physics. We are thus justified in ignoring the interference terms. The computation of the moments involve integrals of Gaussians multiplied by polynomials in $v$. These integrals can be evaluated analytically using a variety of techniques. The results lead to straightforward analytic expressions for the mean,
\begin{equation}
  \mean{v} \approx \sqrt{\frac 2  \pi} e^{-\omega_0^2 t^2/ 2 \sigma_v^2 } \sigma_v + \omega_0 t\, \text{erf}\lf( \frac{ \omega_0 t }{ \sqrt 2 \sigma_v }  \rt)\,,
\end{equation}
and variance,
\begin{align}
  \text{Var}(v)^2 &\equiv \mean{v^2} - \mean{v}^2 \nonumber \\
                &= \sigma_v^2 + \omega_0^2 t^2 - \mean{v}^2\,, \label{eq:ds var}
\end{align}
of $v$, where we have defined
\begin{equation}
  \sigma_v(t) \equiv \frac{ \sqrt{ 1 + 4 \sigma_\omega^4 t^2 } }{ 2 \sigma_\omega }\,.
\end{equation}
These expressions can be compared with exact results obtained from explicit numerical integrations performed on the exact wavefunction given by \eqref{dssolution}.  Excellent agreement is achieved for modest values of $\omega_0/\sigma_\omega$.\footnote{Quantitatively, we see less than 2\% error for $\omega/\sigma_\omega \geq 5$.} 

\begin{figure}
	\includegraphics[width=\linewidth]{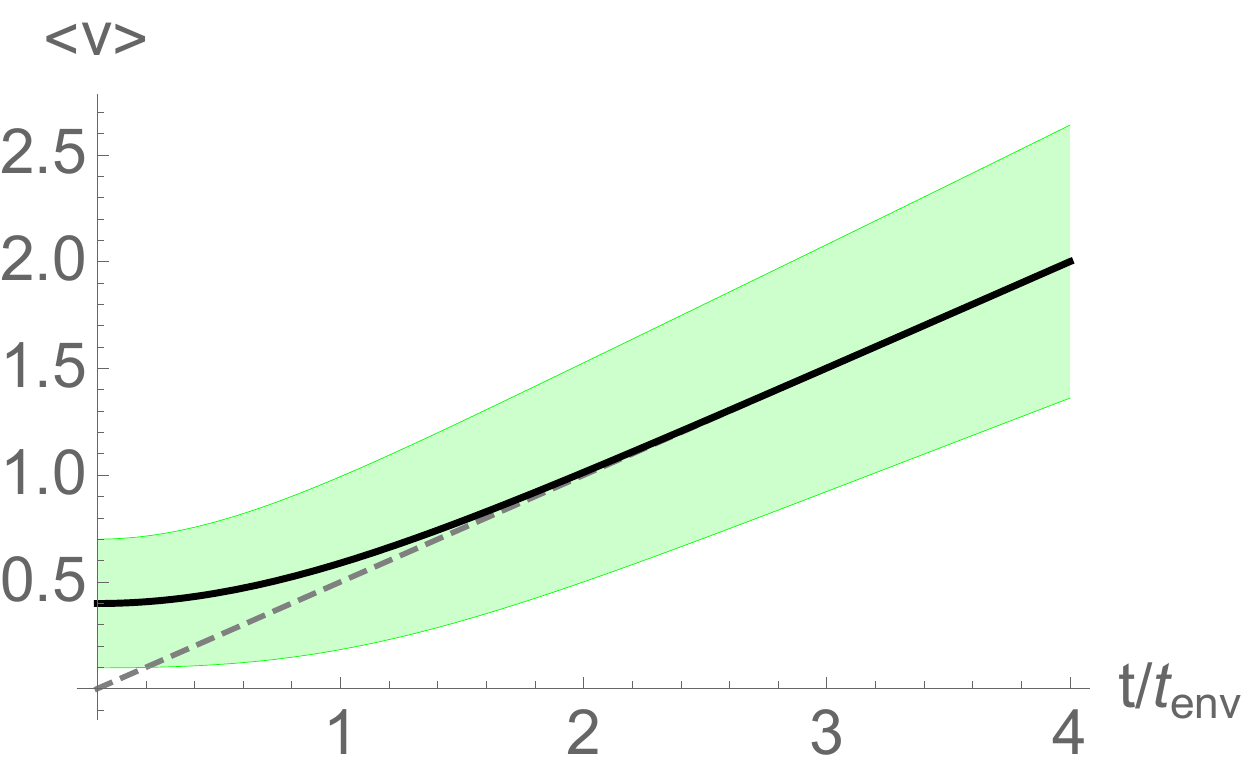}
	\caption{The time evolution of $\mean{v}$ (solid) as compared with the classical solution (dashed). The computed variance is illustrated through a $1 \sigma$ confidence band. ($t_\text{env} \equiv 1/2 \sigma_\omega \omega_0$.)\label{fig:ds_veff}}
\end{figure}

\begin{figure}
	\includegraphics[width=\linewidth]{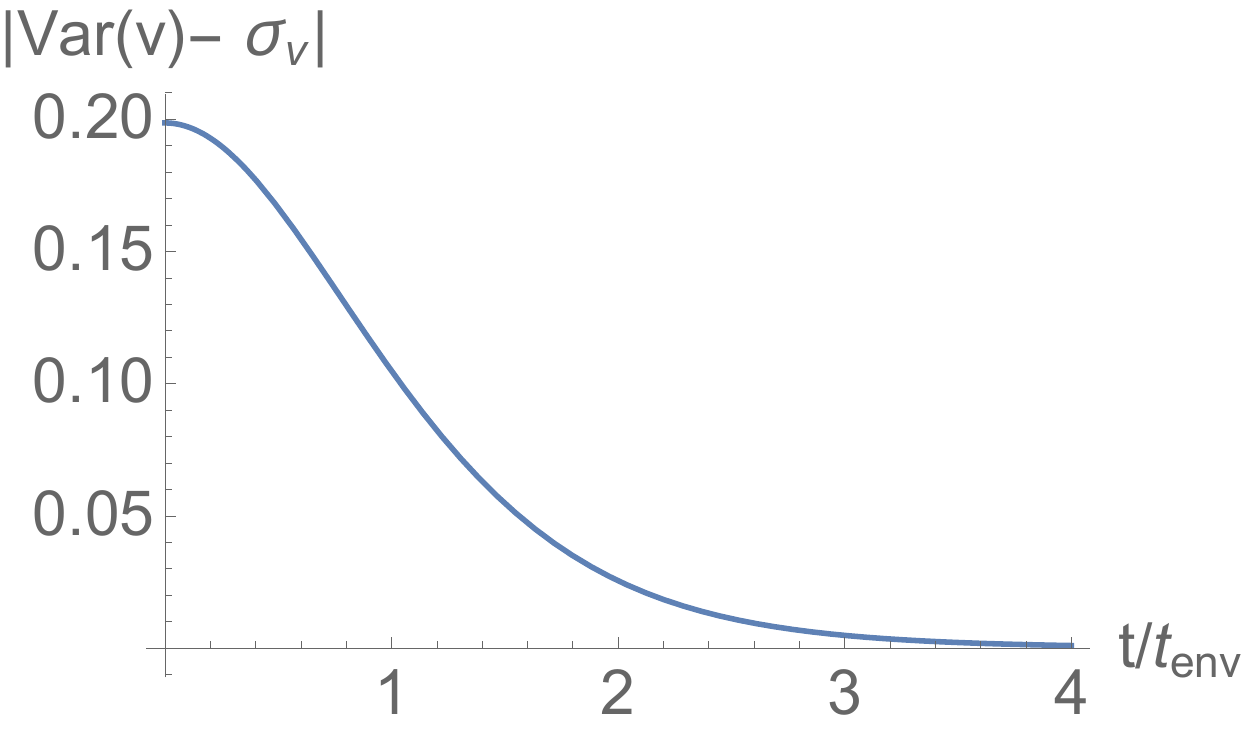}
	\caption{the non-Gaussianities build up near the bounce. This is effect is illustrated via the deviation of the variance from its Gaussian value -- i.e., $|\text{Var}(v)- \mean{v}|$ -- which takes its maximum value at $t=0$ and drops to zero for large $t$. ($t_\text{env} \equiv 1/2 \sigma_\omega \omega_0$.)\label{fig:ds_non_Gauss}}
\end{figure}

The prominent features of $\mean v$ include a deviation away from the classical trajectory on length scales set by $v_\text{env}$. This deviation is  towards the minimum value given by
\begin{equation}
  \mean{v}\big\rvert_{t = 0} \equiv  v_\text{min} = \frac{1}{\sqrt{2\pi}\sigma_\omega}\,.
\end{equation}
Non-Gaussianities can be quantified in terms of the difference between Var$(v)$ and $\sigma_v$. According to \eqref{eq:ds var}, this is given precisely by the departure of $\mean{v}$ from its classical value $\omega_0 |t|$. As was just shown, this departure is negligible when the wavefunction is peaked on length scales larger than $v_\text{env}$ but grows near the bounce achieving a maximum at the bounce time. The bounce can, therefore, be regarded as a quantum process involving a departure from classical behaviour due to interactions between $\mean{v}$ and higher order moments of the wavefunction in $v$ near the bounce region $v \sim v_\text{env}$. The general behaviour of these solutions is illustrated in FIG.~\ref{fig:ds_veff} and FIG.~\ref{fig:ds_non_Gauss}.

\subsection{Numerical Solutions}
\label{sub:numerical solution}

\subsubsection{Specification and Justification of Numerical Methods Used}

In this section, we study the general features of the solutions of our model in the parameter ranges
\begin{equation}\label{eq:param range}
  6 \leq \frac{\omega_0}{\sigma_\omega} \ll \frac{|k_0|}{\hbar}\,
\end{equation}
motivated in \S\ref{sub:generic sa ext} and throughout the text. As noted in \S\ref{sub:generic sa ext}, the parameter range \eqref{eq:param range} is potentially physically relevant because it is a limit in which the bounce can occur well-below the Planck scale and because the behaviour of the self-adjoint extensions becomes universal. Additionally, this limit allows us to approximate the norm of $\Lambda$-eigenstates, as given in \eqref{eq:Psi full}, in an $\omega$-independent way since the real part of the argument of the hyperbolic cosine will dominate its behaviour. Concretely, we have that
\begin{equation}
  \cosh \lf( \frac{\pi k}{2\hbar} + i \frac k \hbar \log \frac \omega \omega_0  \rt) \approx \cosh \frac{\pi k}{2\hbar}\,,
\end{equation}
which grows exponentially in $k$.

This simplification allows us to perform the $\omega$-integrals in \eqref{eq:Psi full} analytically provided we approximate the Gaussian functions used for $E(\omega)$ with the function defined in \eqref{eq:E approx}. Using this and the standard Gaussian for $C(k)$ defined in \eqref{eq:k and E wavefunction}, we obtain
\begin{widetext}
\begin{multline}
  \Psi(v, \varphi, t) = N \int_{-\infty}^\infty \frac{\de k}{\hbar} \int_0^\infty \frac{\de \omega}\hbar \, \text{sech}\lf( \frac {\pi k}{2\hbar} \rt) \exp \lf\{ - \frac{ (k - k_0)^2 }{4\sigma_k^2} + i \frac {k\varphi}{\hbar}  \rt\} \\
      \times e^{-a^2 \omega^2/\hbar^2} \text{Re} \lf[ \lf( \frac {\hbar}{\omega_0} \rt)^{-ik/\hbar} \lf( \frac \omega {\hbar} \rt)^{m - ik/\hbar - 1} \mathcal J_{ik/\hbar} \lf( \tfrac{\omega v}\hbar \rt)\rt]\,,
\end{multline}
where $m$ and $a$ are given in \eqref{eq:ds parameters}\footnote{By inserting factors of $\hbar$ when appropriate.} and
\begin{equation}
  N =  \lf(\frac{\hbar^2 }{(2\pi)^2 \sigma_\omega \sigma_k }\rt)^{1/2} e^{\frac{\omega_0^2}{8 \sigma_\omega^2}} \lf( \frac{\omega_0}\hbar \rt)^{3/2-m}\,.
\end{equation}
In this form, the integrations can be evaluated using the integral
\begin{equation}\label{eq:bessel int}
  \int_0^\infty \de t\, t^{\mu - 1} e^{- a^2 t^2} \mathcal J_{\nu} (vt) = \frac{ \lf( \frac v {2a} \rt)^\nu  \Gamma \lf(  \frac{\nu + \mu}2 \rt) }{2 a^\mu \Gamma(\nu + 1) }  \exp \lf( - \frac {v^2}{4a^2}\rt) \tensor[_1]{F}{_1}\lf( \frac{\nu - \mu}2 + 1, \nu + 1, \frac{v^2}{4a^2} \rt)\,,
\end{equation}
for $\text{Re}(\mu + \nu) > 0$, $\text{Re}(a^2) > 0$, and where $\tensor[_1]{F}{_1}$ is the confluent hypergeometric function. The result is:
\begin{multline}\label{eq:psi analytic}
    \Psi(v, \varphi, t) = \frac {N\Gamma\lf( \frac m 2 \rt) e^{-v^2/4a^2} } {4a^{m} } \int_{-\infty}^\infty \frac{\de k}{\hbar} \text{sech}\lf( \frac {\pi k}{2\hbar} \rt) \exp \lf\{ - \frac{ (k - k_0)^2 }{4\sigma_k^2} + i \frac {k\varphi}{\hbar}  \rt\} \\ \lf[ \frac{ \lf( \frac{\omega_0 v}{2\hbar} \rt)^{ik/\hbar} }{ \Gamma\lf(\tfrac {ik}\hbar + 1\rt) } \tensor[_1]{F}{_1}\lf( 1 - \tfrac m 2 + \tfrac{ik}\hbar, \tfrac {ik}\hbar + 1, \tfrac{v^2}{4a^2}  \rt) + (k \to -k) \rt]\,,
\end{multline}\end{widetext}
where the symbol $(k \to -k)$ implies repeating the first term with $k$ inverted.

Evaluation of the Fourier transform \eqref{eq:psi analytic} is not feasible (or enlightening) in terms of well-known analytic functions. Before proceeding with a numerical evaluation, it is instructive to consider whether the useful approximation employed in the de~Sitter limit may be relevant here. For vanishing $k_0$, we noticed that the condition $\omega_0/\sigma_\omega \gg 1$ implied a regime where the wavefunction only has significant support when $v\omega/\hbar \sim \omega_0/\sigma_\omega \gg 1$. The asymptotic expansion of the Bessel functions for large $v\omega/\hbar$ then lead to an analytically tractable limit. Unfortunately, close inspection of the correction terms in the asymptotic expansion \eqref{eq:asympt} (which can be found explicitly in \cite[\S 10.17]{NIST:DLMF}) reveals that, for non-zero $k$, the correction terms appear as a power series in $k/v\omega$ starting at order 2. Thus, because we are simultaneously taking a limit where $k/\hbar \gg \omega_0/\sigma_\omega$ and $\omega_0/\sigma_\omega \gg 1$, the condition $v \omega/\hbar \sim \omega_0/\sigma_\omega$ is not strong enough to ensure that the higher order corrections in $\hbar/v\omega$ are small. In fact, the dominance of $k$ over $v\omega$ implies that this particular series representation does not converge at all. The analytic tools used in the de~Sitter limit are therefore no longer available here. We thus proceed to evaluate the solutions numerically. 

The $k$-integrations can be performed in a numerically efficient manner by making use of the Fast Fourier Transform (FFT) algorithms. To apply these techniques, we must truncate the integral \eqref{eq:psi analytic} in an appropriate way and choose an optimal lattice spacing or, equivalently, a sampling rate for the Riemann sum. Fortunately, the semi-classicality requirement specifies the narrow range in $k$-space where the wavefunction has support. Moreover, well-known results from sampling theory can be used to specify the optimal sampling rate for computation of the Fourier transform. The details of this procedure, including a specification of the choice of cutoff and sampling rates, is given in Appendix~\ref{sub:fast_fourier_transform_}.

\subsubsection{Results} 
\label{sub:results}

\begin{figure}
  \subfloat[$t=0$ (bounce)]{
    \includegraphics[width=0.5\linewidth]{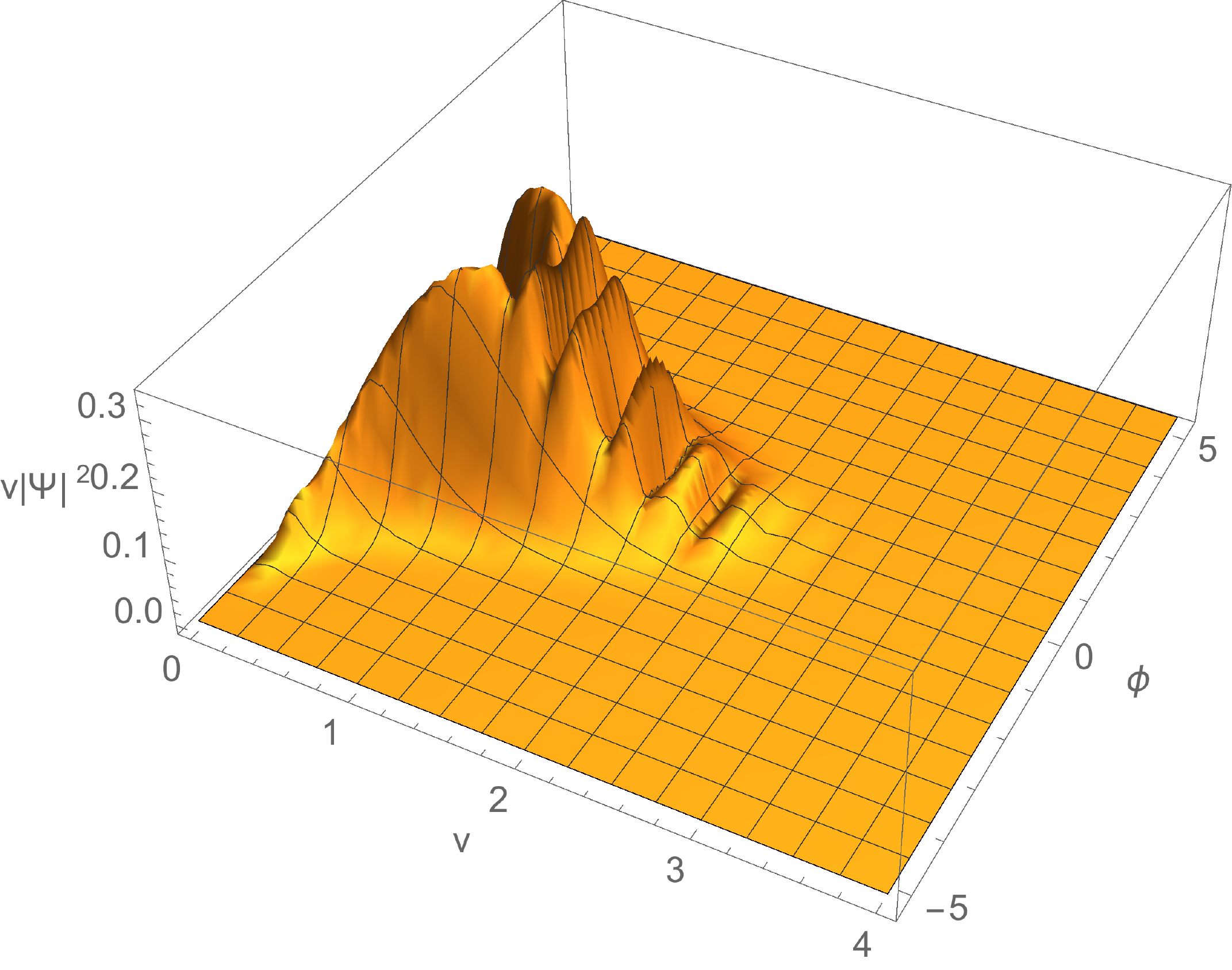}
    \label{fig:bounce animation 4}
  }
  \subfloat[$t=\frac 1 3 t_s$]{
    \includegraphics[width=0.5\linewidth]{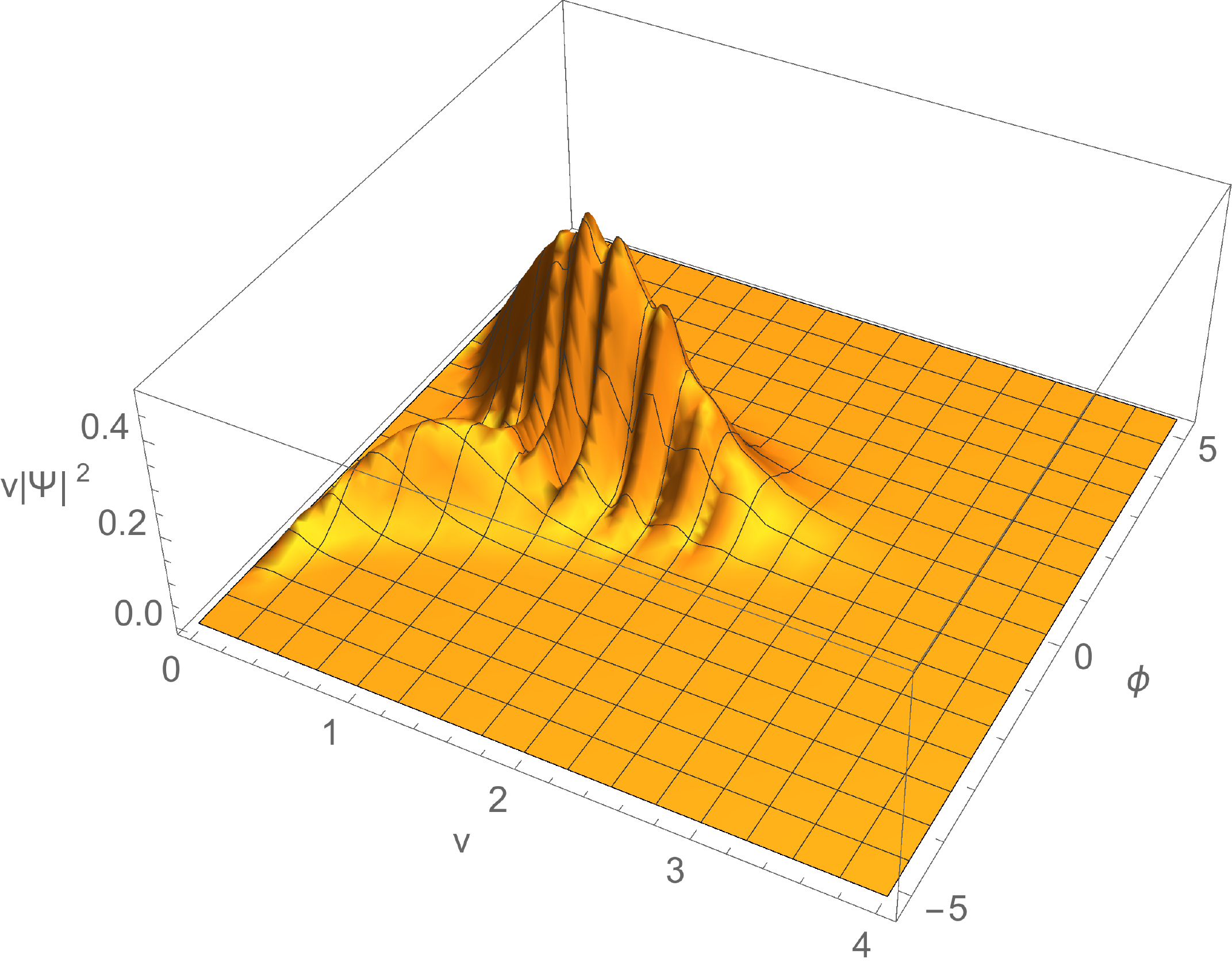}
    \label{fig:bounce animation 3}
  }\\
  \subfloat[$t=\frac 2 3 t_s$]{
    \includegraphics[width=0.5\linewidth]{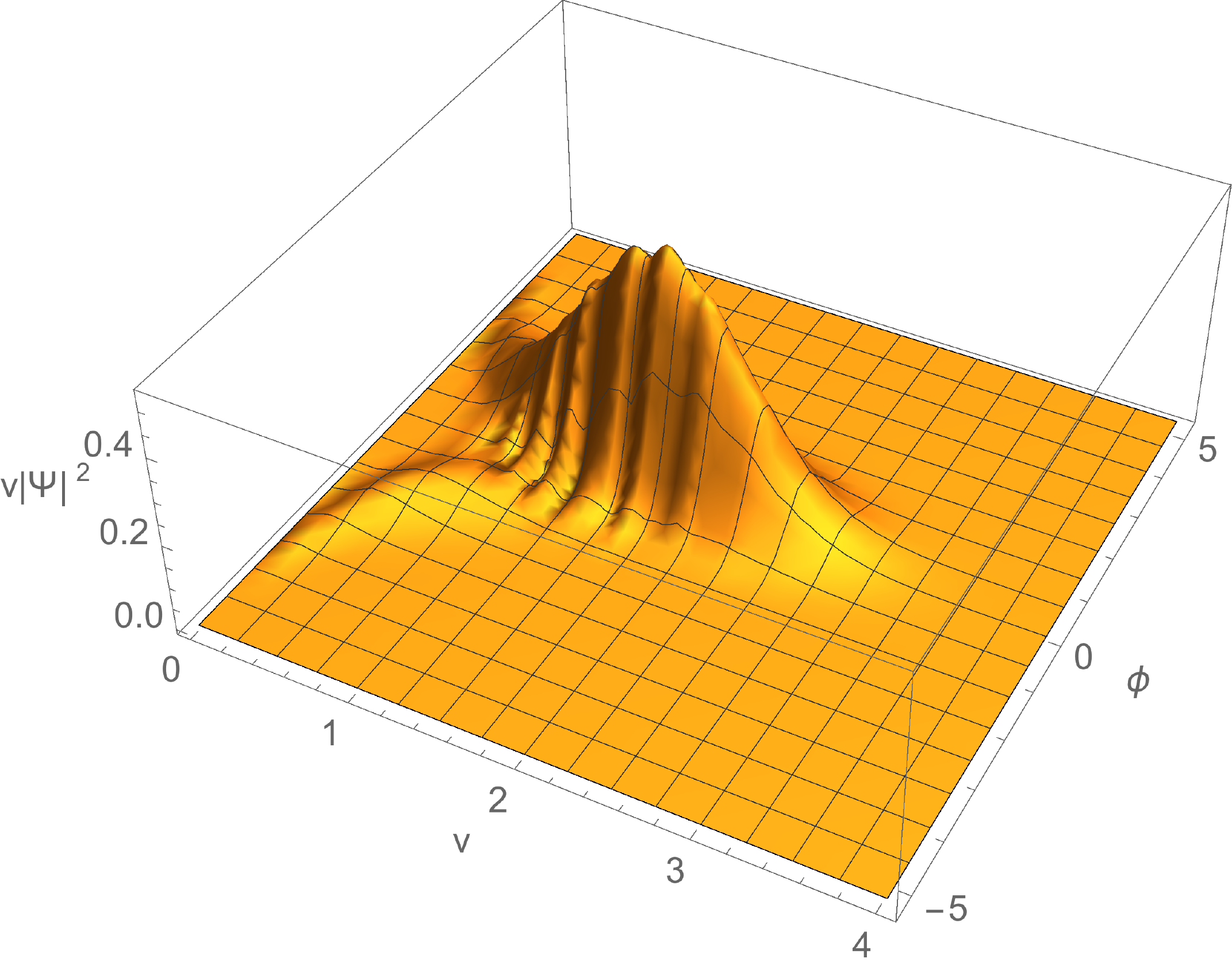}
    \label{fig:bounce animation 2}
  }
	\subfloat[$t=t_s$ (late time)]{
    \includegraphics[width=0.5\linewidth]{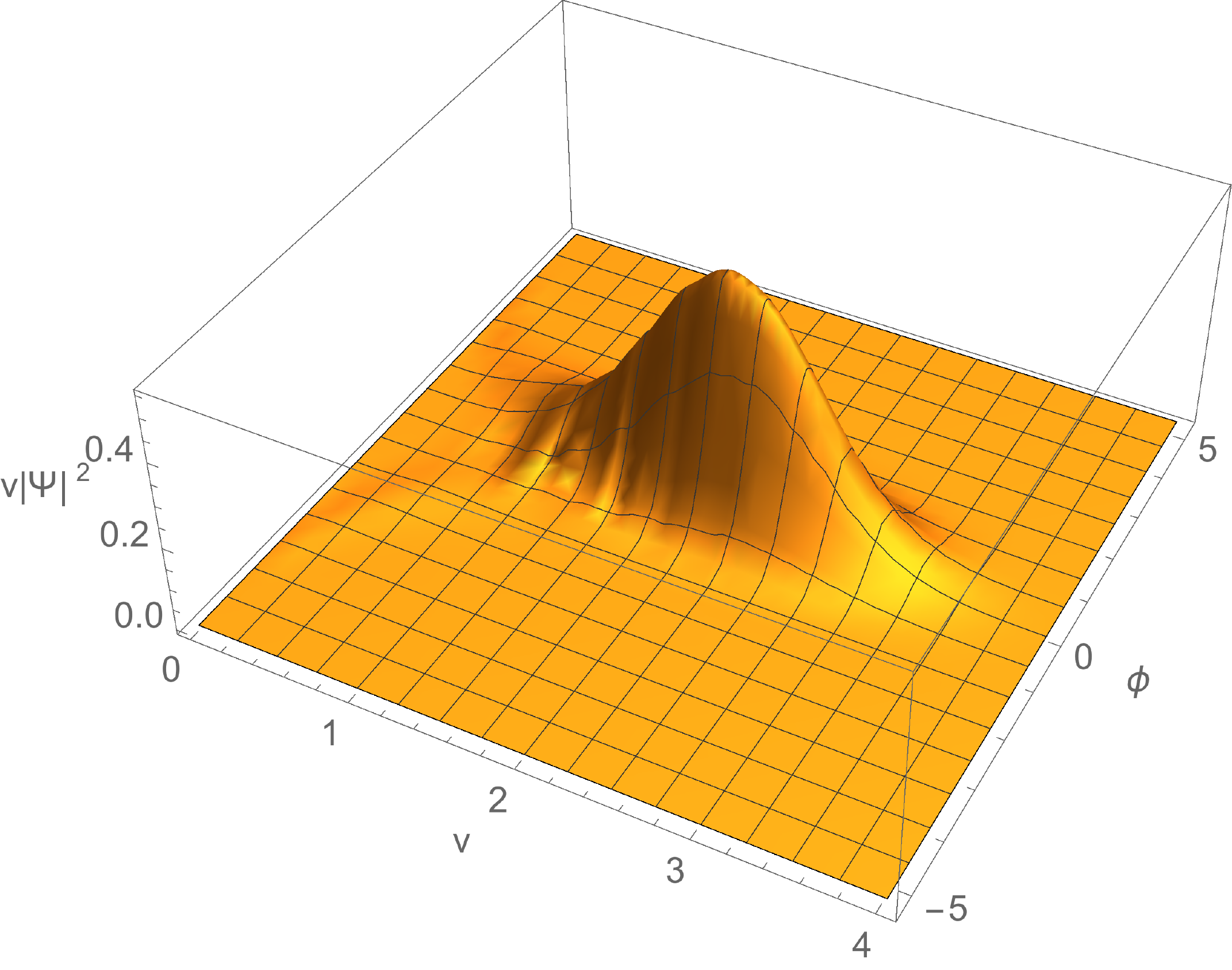}
    \label{fig:bounce animation 1}
  }
  \caption{\label{fig:bounce animation} Animation of the time evolution of the Born amplitude of the wavefunction for $\omega_0/\sigma_\omega = 10$, $k_0/\sigma_k = 10$, and $k_0/\hbar = 10$. ($t_s = \pi \hbar^3/\omega_0^2$). Behaviour is time symmetric about $t=0$.}
\end{figure}

To get a sense of the general behaviour of the solutions, it is useful to plot the time evolution of the wavefunction amplitude from late to early times for modest parameter values. This is illustrated in FIG.~\ref{fig:bounce animation}. The numerical plots confirm visually that certain key features of the dS limit persist when $k$ is non-zero. In particular, we notice that the solutions are characterised by two qualitatively different phenomena: two reasonably localized envelopes representing in- and out-going branches of the wavefunction and rapid beats where these envelopes overlap. This confirms that the qualitative behaviour expected in the Rayleigh scattering limit persists in the full model under this choice of parameters.

Inspection of the exact wavefunction, \eqref{eq:psi analytic}, reveals that the model, under the constraints specified above, depends on the three independent parameters:
\begin{itemize}
  \item [] $\omega_0/\sigma_\omega$: the size of $\omega_0$, which is proportional to the square root of $\Lambda$, in units of the variance in $\omega$;
  \item [] $k_0/\sigma_k$: the size of the scalar field momentum in units of its variance; and
  \item [] $k_0/\hbar$: the size of the scalar field momentum Planck units.
\end{itemize}
We consider the effects of varying each of these parameters individually below.

\begin{figure}
  \includegraphics[width=\linewidth]{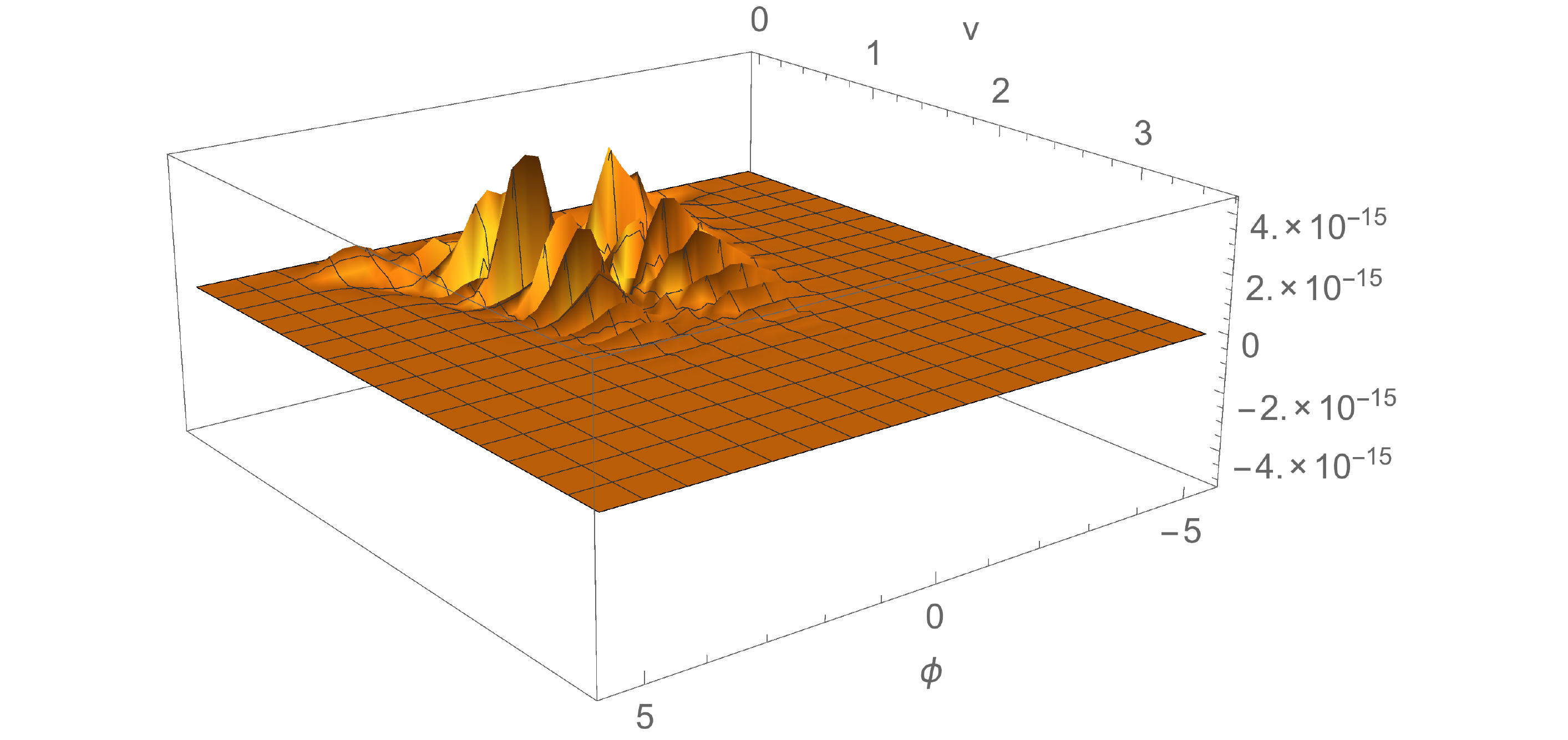}
  \caption{\label{fig:psi_sq_diff} The relative difference between wavefunction amplitudes as $\hbar$ is varied from $1$ to $10$. Differences are barely larger than machine sized but closely follow the beat structure. ($\omega_0/\sigma_\omega = k_0/\sigma_k = 10$ and $t = 0$)}
\end{figure}

The parameter $k_0/\hbar$ encodes the only genuine dependence of the solutions on the value of $\hbar$. We thus expect $k_0/\hbar$ to encode Planck-scale physics. Asymptotically, $k_0/\hbar$ affects the phase shift, $\Delta$, between in- and out-going $\Lambda$-eigenstates according to \eqref{eq:phase}. It is, therefore, reasonable to assume that $k_0/\hbar$ should have a direct effect on the beat physics, which we have already extensively argued to be associated with Planck-scale effects. However, because the limit \eqref{eq:param range} implies both universality of the self-adjoint extension physics and insensitivity of the envelope physics to the physics of the beats, we expect that varying $k_0/\hbar$ while keeping $k_0/\sigma_k$ fixed should have very little impact on our solutions when we restrict our parameters roughly to the regime defined by \eqref{eq:param range}. This can be verified explicitly for modest parameter values. Concretely, if $\omega_0/\sigma_\omega = k_0/\sigma_k = 10$, then we find that changing $\hbar$ from $1$ to $10$ (or, equivalently, varying $k_0/\hbar$ by a factor of 10)  only changes the wavefunction amplitude, when $t=0$, by $4$ parts in $10^{15}$, which is only mildly above machine precision. The largest errors occur precisely where the beat effects are the greatest, reinforcing the hypothesis that the beat physics are genuine Planck effects related to the size of $\hbar$ relative to $k_0$. This is illustrated in FIG.~\ref{fig:psi_sq_diff}.

The remaining two parameters can be varied by fixing $\sigma_\omega = \sigma_k = \hbar = 1$ and then by varying $\omega_0$ and $k_0$ independently. Recall that, in the classical theory, rescaling $v$ by the parameter $s = |k_0|/\omega_0$ allowed us to express the universal behaviour of the classical theory conveniently in terms of \eqref{eq:class ri curve} (see FIG.~\ref{fig:rep inv sols}). Moreover, for our choice of $\sigma_\omega$, the limit \eqref{eq:inflation limit} suggests that the quantity $s$ (in these units) is a particularly convenient parameter for studying the features of the quantum theory. Noting that the de~Sitter limit (i.e., $s=0$) was fully described by $\omega_0/\sigma_\omega$, we choose $s$ and $\omega_0$ (in these units) as relevant parameters.

\begin{figure}
  \subfloat[$v |\Psi|^2$ for $\omega_0/\sigma_\omega = 10$, $s =1$]{
    \includegraphics[width=0.5\linewidth]{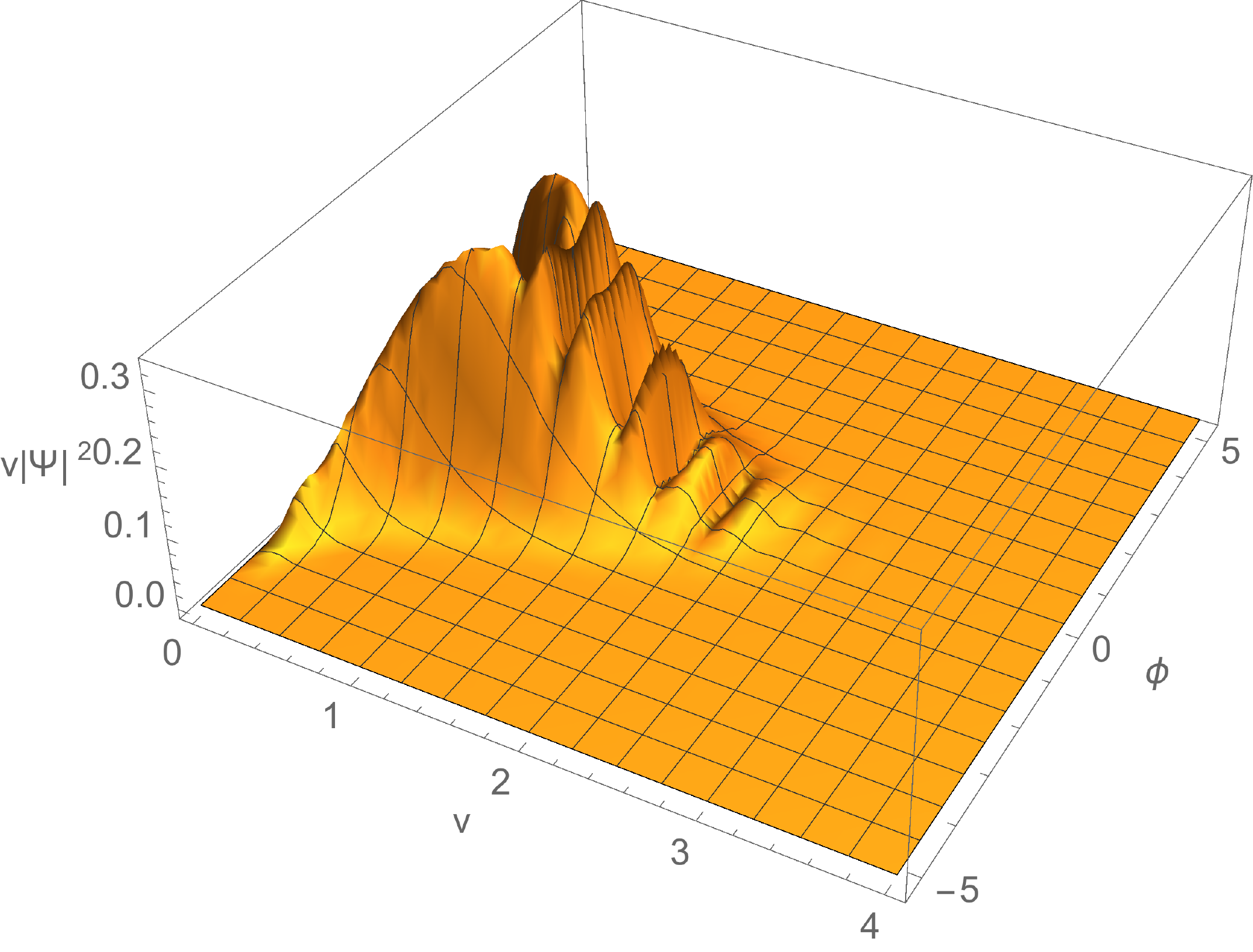}
    \label{fig:bounce_omega_10}
  }\\
  \subfloat[$v |\Psi|^2$ for $\omega_0/\sigma_\omega = 15$, $s=1$]{
    \includegraphics[width=0.5\linewidth]{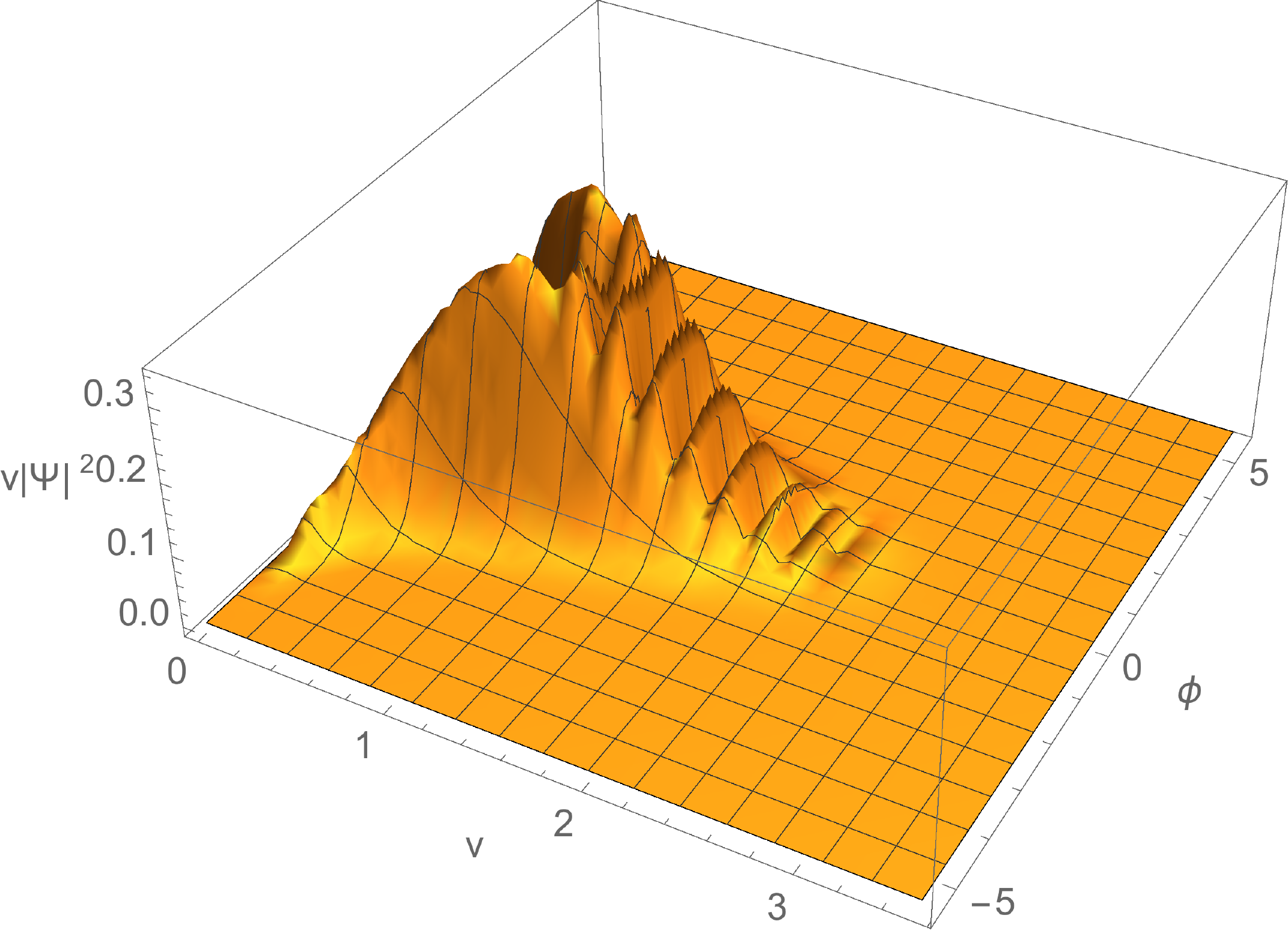}
    \label{fig:bounce_omega_15}
  }
  \subfloat[$v |\Psi|^2$ for $\omega_0/\sigma_\omega = 10$, $s=2$]{
    \includegraphics[width=0.5\linewidth]{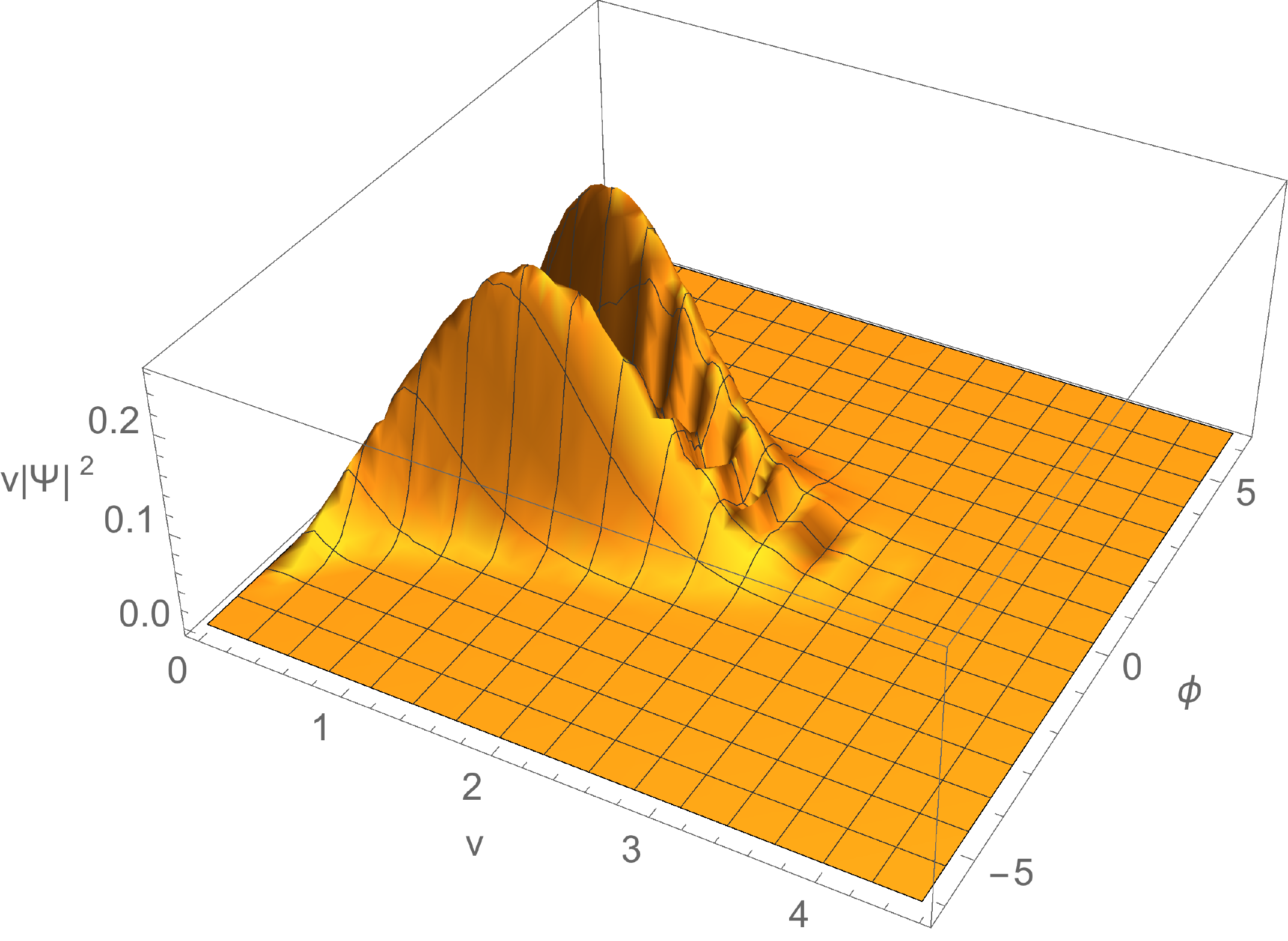}
    \label{fig:bounce_s_2}
  }
  \caption{\label{fig:bounce_comp} Comparison of Born amplitude at bounce time for different choices of $\omega_0$ and $s$ (for $\sigma_\omega = \sigma_k= \hbar = 1$). The beat physics is affected in the same way by $\omega_0$ as it was in the de~Sitter limit (see FIG.~\ref{fig:psi_sq_ds}). Increasing $s$ causes the wavefunction to follow more tightly the classical solutions resulting in less overlap between in- and out-going solutions and, therefore, less prominent beat physics.}
\end{figure}

To study the effects of varying these parameters, we consider two useful plots. The first, inspired by the de~Sitter limit, is the wavefunction amplitude at $t=0$ as one varies $\omega_0$ and $s$. In accordance with our expectations, varying $\omega_0$ has the effect of proportionately varying the beat frequency. The beat frequency itself can be seen to agree, within the resolution of these plots, with the beat frequency in the de~Sitter limit for the same values of $\omega_0$ irrespective of the value of $s$. Conversely, varying $s$ has little effect on the beat frequency. Rather, higher values of $s$ appear to herd the wavefunction more tightly along the classical in- and out-going solutions resulting in less overlap and, consequently, less prominent beat physics. These effects are illustrated in FIG.~\ref{fig:bounce_comp}.

\begin{figure}
  \includegraphics[width=.7\linewidth]{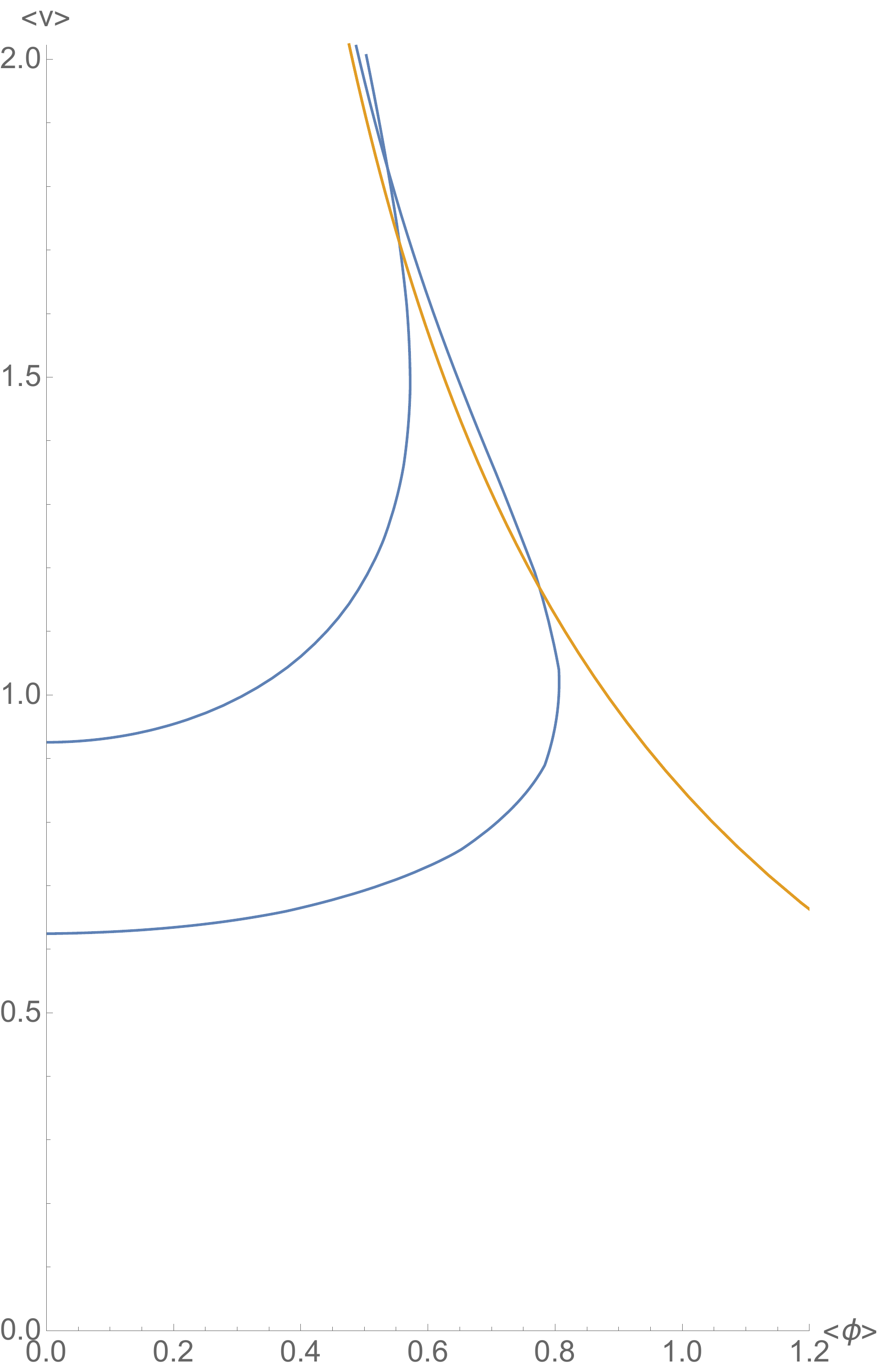}
  \caption{\label{fig:mean v mean phi} A comparison of $\mean{\hat v}/s$ versus $\mean{\hat\varphi}$ for different values of $s$. The top blue line represents $s=1$, the bottom blue line $s=2$, and the yellow line is the classical curve with $\omega_0 = 10$. Increasing $s$ can be seen to decrease $v_\text{min}$ and increase $\varphi_\text{max}$. Changing $\omega_0$ has negligible effect. The figure is symmetric upon the reflection $\varphi \to -\varphi$, which represents $t \to -t$.}
\end{figure}

To quantify more precisely the effects of varying $s$, it is best to plot the numerically computed expectation values of $v$ parametrically against those of $\varphi$, where $v$ is measured in units of $s$. These plots can be compared with the parameter-independent curves obtained in the classical theory in FIG.~\ref{fig:rep inv sols}. The end result is for different values of $s$ at fixed $\omega_0$ is given in FIG.~\ref{fig:mean v mean phi}.\footnote{Changing $\omega_0$ can be seen to have a negligible effect on these curves as can be understood from the fact that $\omega_0$ controls in the beat physics and this has little effect on the expectation values, which depend mostly on the physics of the envelope.} These curves display two distinct and noteworthy features:
\begin{itemize}
  \item [] $\mean{v}$ attains a minimum value, $v_\text{min}$ at the bounce time $t=0$.
  \item [] $\mean{\phi}$ attains a maximium value, $\varphi_\text{max}$ in a region where quantum effects due to Rayleigh scattering take over (i.e., when $v \sim 1/\sigma_\omega$).
\end{itemize}
Changing $s$ has the effect of flattening out this curve so that $v_\text{min}/s$ decreases while $\varphi_\text{max}$ increases. In this way, $s$ determines roughly how long the expectation values remain glued to the classical curve. The smaller the value of $s$, the more rapidly the curve departs from the FLRW solution staying more de~Sitter-like. For both curves, the expectation values cross the classical curve as non-Gaussianities persisting in the individual in- and out-going envelopes are balanced by the non-Gaussianties due to do the superposition of in- and out-going solutions. The former can be observed to fade gradually to zero as $|t|$ increases.


\section{Prospectus} 
\label{sec:concl}

\subsection{Inflationary Cosmology} 
\label{sub:inflationary_cosmology}


The Hubble parameter,\footnote{We add a subscript $0$ to avoid notational ambiguity with the Hamiltonian, $H$, defined earlier.} defined as $H_0 \equiv \dot a/ a$ in a proper-time parametrization (where $N=1$), can be related the variables used in this paper using Hamilton's first equation when $N=1$. We find
\begin{equation}
	H_0 = \frac{\kappa}{2a_0} \pi_v\,,
\end{equation}
where $a_0$ is the fiducial value of $a$. If we treat the Hubble parameter as a phase space variable in a particular time parameterization, it is, therefore, possible to identify an effective Hubble expansion $H_\text{eff}$ as being proportional to the expectation value of $\hat\pi_v$ via\footnote{ The fact that $\hat\pi_v$ is not self-adjoint, as discussed extensively in \cite{Gryb:2017a}, could provide an obstruction to this definition. We posit that, provided inflation occurs much below the Planck scale, this subtle issue should be effectively irrelevant. A more rigorous description of inflation in terms of the self-adjoint operator $\hat\pi_\mu$ can always be provided. }
\begin{equation}
	H_\text{eff} \equiv \mean{\hat H_0} \propto \mean{\hat\pi_v}\,.
\end{equation}
A particularly convenient way of modeling inflationary cosmology, \cite{PhysRevD.42.3936,Liddle:1994dx} , is in terms of a Hamilton--Jacobi formalism where the Hubble parameter, $H_0(\phi)$, is thought of as a relational function of the scalar field $\phi$. In this form, the slow-roll parameters can be expressed relationally as
\begin{align}
	\epsilon_H(\phi) &= \frac{ m_\text{Pl}^2 }{4 \pi} \lf( \frac{H_0'(\phi)}{H_0(\phi)} \rt)^2 \\
	\eta_H(\phi) &= \frac{ m_\text{Pl}^2 }{4 \pi} \frac{H_0''(\phi)}{H_0(\phi)}\,,
\end{align}
where primes denote derivatives with respect to $\phi$. The conditions for slow-roll inflation can then be expressed as $\epsilon_H \ll 1$ and $\eta_H \ll 1$. These conditions often provide a useful approximation for analytically relating observable quantities -- such as the CMB power spectrum amplitude and spectral tilt -- to the parameters of a concrete inflationary model.

At an effective level, we can generalize this formalism by treating the effective Hubble parameter, $H_\text{eff}$, as a function of the expectation value of $\hat\phi$. In terms of our variables, this translates into computing the expectation value of $\hat\pi_v$, or $ \bar{\pi}_v \equiv \mean{\hat\pi_v}$, as a function of the expectation value of $\hat\varphi$, or $\bar\varphi \equiv \mean{\hat\varphi}$. This suggests an alternative parameterization of the slow-roll parameters in terms of our variables as
\begin{align}
	\epsilon_v(\bar\varphi) &\equiv \lf(\frac {\bar {\pi}_v'(\bar\varphi)}{\bar {\pi}_v(\bar\varphi)}\rt)^2 \\
	\eta_v(\bar\varphi) &\equiv \frac {\bar {\pi}_v''(\bar\varphi)}{\bar {\pi}_v(\bar\varphi)}\,,
\end{align}
where the slow-roll conditions are $\epsilon_v \ll 1$ and $\eta_v \ll 1$. These conditions identify a regime of effective inflation. They will be satisfied at least once in an expanding branch of any bouncing cosmology of this kind when $\bar\varphi$ reaches its extremum, where the effective dynamics of $\mean{\hat v}$ and $\mean{\hat \varphi}$ are locally indistinguishable from that of a de~Sitter universe.

In our model, the curve $\bar v(\bar\varphi)$ of FIG.~\ref{fig:mean v mean phi} displays a turnaround point in $\bar\varphi$ shortly after the bounce. At at this turnaround point $\bar\varphi$ evolution becomes momentarily frozen as the Universe is expanding mimicking what happens in a de~Sitter-like epoch of inflation. However, because our model allows for non-trivial superpositions of $\Lambda$, it is possible to obtain this limit of inflation in the Rayleigh-scattering regime where the effective physics of the bounce is taking place at energies far removed from the Planck-scale. This leaves open the possibility that the inflationary regime obtained in our model could be interpreted as an effective epoch of genuine inflationary cosmology.\footnote{Of course, to obtain the observed value of roughly $60$ e-folds of inflation, the free parameters of our model will have to be set to very different values from the ones used to compute the plots of FIG.~\ref{fig:mean v mean phi}.} As our analytic methods are breaking down in precisely the regime where $\bar\varphi$ reaches its maximum value, numerical techniques are required to relate the parameters of our model to the relational slow-roll parameters just defined. Perturbative QFT methods for fully quantum homogeneous backgrounds, such as those used in \emph{hybrid quantization} \cite{FernandezMendez:2012vi}, could be utilized to relate the measurable features of the CMB power spectrum to the novel features of our model. Within such a framework a variety of potentially interesting scenarios could be investigated.\footnote{Such as investigating potential traces of primordial black holes inspired by non-commutative geometry \cite{arraut:2009}.} These will be the subject of future investigations.

\subsection{Reduced-symmetry Models} 
\label{sub:non_symmetric_models}


Our model can be generalized to allow inhomogeneities and anisotropies. Quantized perturbations can be added, in the simplest case, by treating the effective equations for $\bar v$ and $\bar\varphi$ as those of a classical background. Standard techniques have been developed for this purpose \cite{Agullo:2012fc,FernandezMendez:2012vi} and, though highly involved, can be straightforwardly applied to our model using, for example, the (epistemically humble) initial conditions suggested in \cite{Ashtekar:2016pqn}.

The general features of our quantization can also be applied to the unitary quantization of anisotropic Bianchi models \cite{wainwright_ellis_1997}. The Bianchi I Hamiltonian, for example, is trivially accommodated into our framework via a reduced Hamiltonian where the momenta of the anisotropies can be modeled as an effective value of $k$ with the Bessel equations unmodified. Bianchi IX models modify the Bessel equations, but the asymptotic behaviour of the wavefunction near the singularity and near the late-time attractors (i.e., the large $v$ limit) is identical. Since the construction of the self-adjoint extensions depends on the behaviour of the wavefunction near $v=0$ and since the existence of the semi-classical approximation depends on the Gaussianity of the wavefunction near the late-time attractors, one may expect that many of the qualitative features of our model to carry forward to the self-adjoint, semi-classical solutions of the Bianchi IX model. The Bianchi IX model may be particularly valuable for studying general singularity resolution in quantized GR in light of the BKL conjecture \cite{Belinsky:1970ew}.

\section*{Funding}

We are very grateful for the support from the Institute for Advanced Studies and the School of Arts at the University of Bristol and to the Arts and Humanities Research Council. S.G. would like to acknowledge support from the Netherlands Organisation for Scientific Research (NWO) (Project No. 620.01.784) and Radboud University. K.T. would like to thank the Alexander von Humboldt Foundation and the Munich Center for Mathematical Philosophy (Ludwig-Maximilians-Universit\"{a}t M\"{u}nchen) for supporting the early stages of work on this project.

\begin{acknowledgments}
   We are appreciative to audiences in Bristol, Berlin, Geneva, Harvard, Hannover, Nottingham and the Perimeter Institute for comments.
\end{acknowledgments}

\appendix

\section{Regularization of Gaussian} 
\label{sub:regularization_of_gaussian}

In this section, we show that the function \eqref{eq:E approx} provides a useful approximation to a Gaussian in the limit \eqref{eq:moment limit}. This approximation achieves two goals: i) it converges quickly and efficiently to a Gaussian in the desired limit, and ii) it can be integrated over Bessel functions analytically (using the integral \eqref{eq:bessel int}).

The desired approximation is of the general form:
\begin{equation}\label{eq:Gauss reg}
  e^{- \sigma_x^2 (p- p_0)^2} \approx N p^n e^{- c^2 p^2}\,,
\end{equation}
where $n \in \mathbbm R$, in the limit
\begin{equation}
  \sqrt {n} \gg 1\,.
\end{equation}
As we will see, for practical purposes, $\sqrt {n} > 10$ gives an adequate approximation. We wish to fix the parameters $n$, $c$, and $N$ in terms of $p_0$ and $\sigma_x$.

Note that the maximum of \eqref{eq:Gauss reg} obtains when
\begin{equation}
p_\text{max} = \sqrt{\frac{n}{2 c^2}} \to p_0\,.
\end{equation}
We therefore wish to show that this occurs in the limit $p_0 \gg 1/\sigma_x$, which is the general form required by our semi-classical considerations. To this end, we can now fix $N$ and $c$ by matching successive terms in the Taylor series expansion of both sides of \eqref{eq:Gauss reg} about $p = p_0$, on the LHS, and $p = p_\text{max}$, on the RHS. The first term tells us that
\begin{equation}
  N = \lf( \frac {2 e c^2}{n} \rt)^{n/2}\,.
\end{equation}
The second term is satisfied if we set $p_0 = p_\text{max}$ and the quadratic term, after a little algebra, relates the inverse width to $c$ via
\begin{equation}
  c = \frac{\sigma_x}{\sqrt 2}\,,
\end{equation}
for $c>0$ and $\sigma_x >0$. Putting this all together gives
\begin{align}
  e^{- \sigma_x^2 (p- p_0)^2} &\approx \lf(\frac p {p_0}\rt)^{\sigma_x^2 p_0^2
} e^{- \frac{\sigma_x^2 p_0^2}2 \lf( (p/p_0)^2 - 1 \rt)} \nonumber \\ &= \exp\lf[ - \frac{\sigma_x^2 p_0^2}2 \lf( \lf(\tfrac p {p_0}\rt)^2 - \lf(1 + 2 \log(\tfrac p {p_0}) \rt) \rt)  \rt]\,.\label{eq:Gauss approx}
\end{align}

It remains to estimate the error in this approximation. A simple way to quantify this error is to compare the integral of both sides of \eqref{eq:Gauss approx} over $p \in [0,\infty)$ (assuming $p_0\gg 1/\sigma_x$). For the LHS, we have
\begin{align}
  I &\equiv \int_0^\infty \de p  \lf(\frac p {p_0}\rt)^{\sigma_x^2 p_0^2
} e^{- \frac{\sigma_x^2 p_0^2}2 \lf( (p/p_0)^2 - 1 \rt)} \nonumber\\ &= \frac {p_0}{2} \lf(\frac{e}{a^2}\rt)^{a^2} \Gamma\lf( a^2 + \tfrac 1 2 \rt)\,,
\end{align}
where $a^2 = \sigma_x^2 p_0^2/2$. Using Sterling's approximation,
\begin{equation}
  \Gamma( z ) = \sqrt{ \frac{2 \pi}{z} } \lf( \frac z e \rt)^z \lf( 1 + \mathcal O(1/z)  \rt)  \,,
\end{equation}
we obtain
\begin{equation}
  \Gamma(a^2 + 1/2) = \sqrt{\frac{2 \pi}{a^2}} \lf( \frac{ a^2 }{e} \rt)^{a^2}\lf( 1 + \mathcal{O}\lf(\frac 1 {a^2} \rt) \rt)
\end{equation}
so that
\begin{equation}
  I = \frac{\sqrt{\pi}}{\sigma_x} + \mathcal{O}\lf(\frac 1 {\sigma_x^2 p_0^2} \rt)\,.
\end{equation}
Since the area under the RHS is $\sqrt{\pi}/\sigma_x$, the error scales like $1/(\sigma_x p_0)^2$, which is vanishingly small illustrating the efficiency of this approximation.

\section{Fast Fourier Transform (FFT) for $k$ integration} 
\label{sub:fast_fourier_transform_}

To perform the $k$-integrations numerically, it is computationally efficient to make use of a Fast Fourier Transform (FFT) algorithm. In order to do this, we first note that the $k$-eigenfunctions are complex exponentials so that integration over them takes the form of a Fourier transform:
\begin{equation}
   F(\varphi) = \frac 1 {\sqrt {2\pi}} \int_{-\infty}^{\infty} \de k\, e^{i k \varphi} f(k)\,.
\end{equation}
We first approximate this integral by imposing the cutoffs $k_\text{min}$ and $k_\text{max}$ respectively. Because of our choice of $C(k)$, the amplitude of our $k$-space integral is localized around $k_0$ with a variance roughly determined by the Gaussian width $\sigma_k$. For practical purposes, therefore, it is sufficient to set $k_\text{min/max} = k_0 \mp 6 \sigma_k$. To convert this to a FFT computation, we then further approximate the truncated integral by a Riemann sum via
\begin{equation}
   F(\varphi) \approx \frac {1} {\sqrt {2\pi}} \sum_{r = 1}^{n} e^{i k_r \varphi} f_r \Delta k\,,
\end{equation}
using a large number $n \gg 1$ of equally spaced intervals $\Delta k = (k_\text{max} - k_\text{min})/(n-1)$ and where
\begin{align}
    k_r &= k_\text{min} + (r-1) \Delta k & f_r &= f(k_r)\,.
\end{align}

Most FFT algorithms will return the Fourier series at equally spaced intervals
\begin{align}
    \varphi_s &\equiv \varphi_\text{min} + (s - 1) \Delta \varphi & \Delta \varphi &\equiv \frac{\varphi_\text{max} -  \varphi_\text{min}}{n-1}\,,
\end{align}
on its image $\in (\varphi_\text{min}, \varphi_\text{min})$. Using these definitions, we can re-arrange the approximate Fourier transform as
\begin{multline}\label{eq:fourier series}
    F(\varphi_s) \approx \frac {e^{i k_\text{min}(\varphi_s - \varphi_\text{min})  }} {\sqrt {2\pi}}   \\ \times \sum_{r = 1}^{n} \exp\lf(  2\pi i \frac{(r-1)(s-1)}n b \rt) e^{i k_r \varphi_\text{min}} f_r \Delta k\,,
\end{multline}
where
\begin{equation}
  b = \frac{ n \Delta k \Delta \varphi}{2 \pi}\,.
\end{equation}
The quantity in brackets is in the standard form for input into an FFT algorithm. For example, in \emph{Mathematica}, the bracketed quantity can be computed using the `Fourier' command of the shifted function
\begin{equation}
  \tilde f_r \equiv e^{i k_r \varphi_\text{min}} f_r
\end{equation}
and with Fourier parameters set to $\pb a b \to \pb 1 b$.

In the above procedure, we have chosen a specified image $(\varphi_\text{min}, \varphi_\text{max})$ and domain $(k_\text{min}, k_\text{max})$ for the Fourier series. It is therefore crucial to choose the sample size, $n$, in a way that will avoid what, in signal processing, is called \emph{aliasing}. By examining the argument of the complex exponential in \eqref{eq:fourier series}, we can see that there is a periodicity in $\varphi$, $T_\varphi$ given by
\begin{equation}
  T_\varphi = \frac {2\pi}{\Delta k}\,.
\end{equation}
To avoid aliasing we must take the minimum number of samples
\begin{equation}
  n_s \equiv 1 + \frac { (k_\text{max} - k_\text{min})(\varphi_\text{max} - \varphi_\text{min}) }{2 \pi}
\end{equation}
or
\begin{equation}
  b = 1 + \frac{2\pi}{ (k_\text{max} - k_\text{min})(\varphi_\text{max} - \varphi_\text{min})  }
\end{equation}
so that $T_\varphi > \varphi_\text{max} - \varphi_\text{min}$, and the periodicity is larger than the range of the image we are interested in. This is equivalent to sampling at the \emph{Nyquist} frequency, $f_s$, above which no additional information is, in principle, gained. In this context, because we have a cutoff on both the image and the domain of the Fourier series, only a finite number $n_s$ of samples are required to specify everything we can know about this function under these restrictions. The Nyquist--Shannon sampling theorem indicates that full knowledge of the function $F(\varphi)$ can be obtained using the optimized sinc-interpolation function\footnote{This can be seen by noting that the Fourier series we are after is the Fourier transform of the original function times a rectangular step function multiplied a Dirac comb. Since the Fourier transform of a Dirac comb is another, shifted, Dirac comb and the Fourier transform of the step function is a sinc function, the desired Fourier series is the result of convolution with both of these functions. }
\begin{equation}
  F(\varphi) = \sum_{r=1}^{n} F(\varphi_n) \text{sinc}\lf( \frac{ \varphi - \varphi_r }{\varphi_\text{max} - \varphi_\text{min}}  \rt)\,.
\end{equation}
However, as this idealized interpolation is rather slow, it is preferable to use a more computationally efficient spine-interpolation method instead for plotting and numerical integration purposes. We thus perform a moderate amount of oversampling by sampling at a rate of $2 n_s$ to accommodate non-optimal interpolation. The errors, therefore, in approximating $F(\varphi)$ are largely dominated by the truncation that has been performed in momentum space. Note that this procedure is extremely computationally efficient since the computational time for the FFT scales like $n_s \log n_s$ -- which translates to $ z^2 \log z $, where $z$ is the number of variances away from the mean to put the cutoff -- while the relative error drops like $e^{-z^2}/z$ (i.e., the asymptotic expansion of the complimentary error-function).


\newpage

\bibliographystyle{utphys}
\bibliography{mach,Masterbib}

\end{document}